\documentstyle[emulateapj]{article}

\lefthead{Nagao, Taniguchi, \& Murayama}
\righthead{The HINER of Seyfert Galaxies}

\begin{document}
\submitted{To Appear in the Astronomical Journal June 2000 Issue}
\title{The High-Ionization Nuclear Emission-Line Region of Seyfert Galaxies}

\author{Tohru NAGAO, Yoshiaki TANIGUCHI, and Takashi MURAYAMA}
\affil{Astronomical Institute, Graduate School of Science, Tohoku University,
       Aramaki, Aoba, Sendai 980-8578, Japan}

\begin{abstract}

Recently Murayama \& Taniguchi proposed that a significant part of
the high-ionization nuclear emission-line region (HINER)
in Seyfert nuclei arises from the inner wall of dusty tori because type 1
Seyfert nuclei (S1s) show the excess HINER emission with respect to type 2
Seyfert ones (S2s).
This means that the radiation from the HINER
has the anisotropic property and thus statistical properties of 
the HINER emission can be used to investigate the viewing angle toward
dusty tori for various types of Seyfert nuclei.
In order to investigate viewing angles toward
narrow-line Seyfert 1 galaxies (NLS1s) and intermediate types of Seyferts
(i.e., type  1.5, 1.8, and 1.9 Seyfert galaxies; hereafter
S1.5, S1.8, and S1.9, respectively), we apply this HINER test to them.
We also apply the same test for S2s with/without the hidden broad line region.
A sample of Seyfert nuclei analyzed here consists of 124 Seyferts
compiled from the literature.

Our main results and suggestions are as follows. 
(1) The NLS1s are viewed from a more face-on view toward dusty tori
than the S2s. However, the HINER properties of the NLS1s are indistinguishable
from those of the S1s.
(2) The S1.5s appear to comprise heterogeneous populations; e.g.,
a) some of them may be seen from an
intermediate viewing angle between S1s and S2s, 
b) some S1.5s are basically S1s but
a significant part of the broad-line region (BLR) emission is 
accidentally obscured
by dense, clumpy gas clouds, or c) some S1.5s are basically
S2s but a part of the BLR emission can be seen from
some optically-thin regions of the dusty torus.
(3) The S1.8s, the S1.9s and the objects showing either
a broad Pa$\beta$ line or polarized broad Balmer lines are seen from a
large inclination angle and the emission from the BLRs of such objects
reaches us through optically-thin parts of dusty tori.
These three results support strongly the current unified model of 
Seyfert nuclei.
And, (4) the line ratios of [Fe {\sc x}]$\lambda$6374
to the low-ionization emission-lines have rather isotropic property
than those of [Fe {\sc vii}]$\lambda$6087.
Therefore it is suggested that the [Fe {\sc x}]$\lambda$6374 emission
is not useful investigating the viewing angle toward the dusty torus in 
Seyfert nuclei. The most plausible reason seems
that the [Fe {\sc x}]$\lambda$6374 emission is spatially 
extended and thus its strength tends to show less viewing angle 
dependence.

\end{abstract}


\section{INTRODUCTION}

Seyfert galaxies have been broadly classified into two classes based on the
presence or absence of broad emission lines in their optical spectra
(Khachikian \& Weedman 1974): Seyfert galaxies with broad lines are type 1
(hereafter S1) while those without broad lines are type 2 (S2).
According to the current unified model of Seyfert nuclei
(Antonucci \& Miller 1985; see for a review Antonucci 1993),
this difference between S1 and S2 can be explained as follows.
The broad-line region (BLR) is located in the very inner region 
(e.g., a typical radial
distance from the central black hole is $r \sim$ 0.01 pc; e.g., Peterson 1993)
and is surrounded by a geometrically and optically thick dusty torus.
Therefore, the visibility of the central engine as well as the BLR
is strongly affected by the viewing angle toward the dusty torus
and then the difference between S1s and S2s is naturally understood.
Indeed, this unified scheme has been supported 
by various observational results,
for example, obscured X-ray emission in S2s (Awaki et al. 1991;
Rush et al. 1996), colors of mid-infrared (MIR) emission 
(Pier \& Krolik 1992, 1993;
Murayama, Mouri, \& Taniguchi 2000),
MIR luminosity distributions (Heckman, Chambers, \& Postman 1992; 
Maiolino et al. 1995),
polarized broad emission lines mentioned below, and the results of
multi wavelength observational tests (Mulchaey et al. 1994).
In order to understand Seyfert nuclei and 
active galactic nuclei (AGNs)
more comprehensively, any new observational tests toward the unified model
are very important.

In addition to the traditional two types of Seyfert nuclei,
it is known that some Seyfert nuclei show intermediate properties
between S1 and S2;
type 1.2 (S1.2), type 1.5 (S1.5), type 1.8 (S1.8), and type 1.9 (S1.9)
(Osterbrock \& Koski 1976; Osterbrock 1977, 1981b;
Cohen 1983; Winkler 1992; Whittle 1992), which show both the narrow
and broad components in the Balmer emission lines.
It is also noted that 
the objects without BLR in their optical spectra (i.e., S2s)
do not comprise a simple population.
First, some S2s show a broad Pa$\beta$ line 
(Goodrich, Veilleux, \& Hill 1994; Hill, Goodrich, \& Depoy 1996;
Veilleux, Goodrich, \& Hill 1997), providing evidence for highly reddened
BLRs in these objects.
Second, the hidden BLR is detected only in a part ($\sim$ 20\%) of S2s 
in the polarized optical spectra (Antonucci \& Miller 1985; 
Miller \& Goodrich 1990; Tran, Miller, \& Kay 1992; 
Kay 1994; Tran 1995a, 1995b, 1995c);
the survey promoted by Lick Observatory found 10 S2s 
with the hidden broad line among 50 S2s.

Another important type of Seyfert nuclei is narrow-line Seyfert 1 
galaxies (NLS1s; Davidson \& Kinman 1978).
Optical emission-line properties of the NLS1s are summarized
as follows (e.g., Osterbrock \& Pogge 1985).
(1) The Balmer lines are only slightly broader than the forbidden
lines such as [O {\sc iii}]$\lambda$5007 (typically less than 
2000 km s$^{-1}$). This property makes NLS1s a distinct type of S1s.
(2) The [O {\sc iii}]$\lambda$5007/H$\beta$ intensity ratio is smaller than 3.
This criterion has introduced to discriminate S1s from S2s by 
Shuder \& Osterbrock (1981). 
(3) They present strong Fe {\sc ii} emission lines 
which are often seen in S1s but generally not in S2s.
And, (4) the soft X-ray spectra of NLS1s are very steep 
(Puchnarewicz et al. 1992;
Boller, Brandt, \& Fink 1996; Wang, Brinkmann, \& Bergeron 1996) and 
highly variable (Boller et al. 1996; Turner et al. 1999a). 
Because of these complex properties, 
it has not yet been fully understood 
what NLS1s are in the context of the current 
unified model of Seyfert nuclei while various models 
for NLS1s have been proposed 
(see for reviews Boller et al. 1996; Taniguchi, Murayama, \& Nagao 1999).

Recently, Murayama \& Taniguchi (1998a; hereafter MT98a) have found that S1s
have excess [Fe {\sc vii}]$\lambda$6087 emission with respect to S2s.
This means that a significantly large fraction of the high-ionization nuclear
emission-line region (HINER; Binette 1985; 
Murayama, Taniguchi, \& Iwasawa 1998)
traced by [Fe {\sc vii}]$\lambda$6087 resides in a viewing-angle dependent
region; i.e., the inner wall of dusty tori
(Murayama \& Taniguchi 1998b; see also Pier \& Voit 1995).
Accordingly, it turns out that the HINER provides the indicator of the
viewing angle for dusty tori of Seyfert nuclei.
In this paper, we report on our statistical analysis of the HINER 
in the various types of Seyfert nuclei.

\section{DATA}

\subsection{Classification of Seyfert Nuclei}

As mentioned in Section 1, there are a number of sub-types of Seyfert
nuclei. Summarizing their definitions and properties,
we broadly re-classify all the objects in the following way
(see Table 1).
1) The type of S1.2 is included in the type of S1.
These Seyferts together with typical S1s are abbreviated as 
BLS1s (broad-line type 1 Seyferts)
because there is another type of S1s; i.e., NLS1s.  
2) The type of S1.5 is kept as a distinct type
because Seyfert galaxies
belonging to this type are more numerous than 
those of the other intermediate-type Seyferts. 
3) The types of S1.8 and S1.9 are basically included into the type of S2.
The BLRs in these galaxies are reddened more seriously than those in both 
BLS1s and S1.5s. In this respect, S2s with the BLR detected only in their
infrared  spectra (e.g., broad Pa$\beta$ emission; hereafter
S2$_{\rm NIR-BLR}$) share the same 
BLR properties. Therefore, we refer these types of Seyferts as
type 2 Seyferts with the reddened BLR (S2$_{\rm RBLR}$).
4) Another important type of S2s is S2s with the hidden BLR
which is detected in optical polarized spectra\footnote{
This type is referred either as S3 (Tran 1995a),
as S1h (V\'{e}ron-Cetty \& V\'{e}ron 1998), 
or as S2$^+$ (Taniguchi \& Anabuki 1999).
}.
In this paper, we refer this type as S2$_{\rm HBLR}$.
5) S2s either with the reddened BLR or with the hidden BLR are also 
referred as S2$^+$; i.e., S2$^+$ = S2$_{\rm RBLR}$ + S2$_{\rm HBLR}$. 
6) In contrast, S2s without any evidence for the BLR are
referred as S2$^-$ following Taniguchi \& Anabuki (1999). 
And, 7) both types of S2$^+$ and S2$^-$ are referred as S2$_{\rm total}$
when necessary; i.e., S2$_{\rm total}$ = S2$^+$ + S2$^-$.
Our classification scheme is summarized in Table 1.

In some cases, the classification is assigned differently
to a certain Seyfert nucleus among the literature.
Therefore, in Table 2,
we compare the types of our sample objects with those given
in some previous papers (Dahari \& De Robertis 1988a; Stephens 1989;
Whittle 1992; Cruz-Gonz\'{a}lez et al. 1994;
V\'{e}ron-Cetty \& V\'{e}ron 1998).
The type adopted in this paper for each galaxy is given in the last 
column of Table 2.
The objects classified as NLS1s
in the previous literature (Osterbrock \& Pogge 1985; Stephens 1989;
Boller et al. 1996; V\'{e}ron-Cetty \& V\'{e}ron 1998;
Vaughan et al. 1999) are categorized as NLS1.

\subsection{Data}

In order to investigate HINER properties of various types of Seyfert galaxies,
we are interested in the following high-ionization emission lines; 
[Fe {\sc vii}]$\lambda$6087, [Fe {\sc x}]$\lambda$6374\footnote{It is
noted that Osterbrock (1977) misidentified
[Fe {\sc x}]$\lambda$6374 as Fe {\sc ii} $\lambda$6369 
(see Osterbrock 1981a).}, and [Fe {\sc xi}]$\lambda$7892.
In addition to these lines, we are also interested in the following 
low-ionization emission lines; [O {\sc iii}]$\lambda$5007, 
[S {\sc ii}]$\lambda\lambda$6717,6731, and [O {\sc i}]$\lambda$6300,
because of the comparison with the high-ionization emission lines.
These lines are simply referred as [Fe {\sc vii}], [Fe {\sc x}],
[Fe {\sc xi}], [O {\sc iii}], [O {\sc i}], and [S {\sc ii}], respectively.
Here we should mention that some fraction of
the [O {\sc iii}] emission arises from the inner wall of 
dusty tori (Pier \& Voit 1995; Murayama \& Taniguchi 1998b).
Therefore, it seems better to use more low-ionization emission lines
such as [O {\sc i}] or [S {\sc ii}] as a normalization emission line.
This is the reason why we have compiled the data of not only [O {\sc iii}]
but also [O {\sc i}] and [S {\sc ii}].
Though [N {\sc ii}]$\lambda$6583 is also one of important
low-ionization emission lines, we do not use this line
because the deblending [N {\sc ii}] from H$\alpha$ may not be well done
if the spectral resolution is not so high.
We have compiled the emission line data from the literature (Table 3)
which are spectroscopic studies at wavelengths covering the emission lines
of our interests.
The number of compiled objects is 227; i.e., 31 NLS1s, 58 S1s, 67 S1.5s, 
31 S2$^+$s, and 40 S2$^-$s. 

The detection rates of [Fe {\sc vii}], [Fe {\sc x}], and [Fe {\sc xi}]
in the sample are given in Table 4.
The fraction of objects with at least one of these high-ionization
lines detected are also given.
This table shows that the detection rate for the S2$^+$ (83.9 \%) 
is higher than those for the other types (38.7 \% $\sim$ 64.2 \%). 
Although the reason for this fact is not clear,
it may be partly because the average redshift of the S2$^+$s is smaller 
than those of the other types (see section 2.3.1) 
and accordingly those objects might be observed with higher S/N.
Here we mention that we do not use any upper limit data in our study.

We choose the objects which show [Fe {\sc vii}] and/or [Fe {\sc x}] from
the sample, and consequently, the object number of our sample is 
124 including 9 
radio-loud galaxies\footnote{In this paper, we define the radio-loud galaxy
as the one which satisfies the criterion of {\it R} $>$ 500, where {\it R}
is the ratio of radio ($\lambda$ = 6 cm) to optical ({\it B} band) flux 
density. Here the {\it R} is defined as follows; the optical flux density 
$S_{\rm opt}$ at {\it B}-band
are calculated from the relation {\it B} = --2.5 $\log S_{\rm opt}$ -- 48.36
(Schmidt \& Green 1983), and {\it R} is derived from dividing the radio flux 
density at the wavelength of 6 cm by this $S_{\rm opt}$.
}; i.e., 12 NLS1s, 23 BLS1s,
43 S1.5s (including 3 radio-loud galaxies), 27 S2$^+$s (including 2 
radio-loud galaxies), and 19 S2$^-$s (including 4 radio-loud galaxies).
Although the sample is not a statistically complete one in any sense,
the data set is the largest one for the study of HINER ever compiled.

In Table 5, the redshift, the apparent {\it B} magnitude, the absolute 
{\it B} magnitude\footnote{In this paper, we adopt a Hubble constant 
{\it H}$_{0}$ = 50 km s$^{-1}$ Mpc$^{-1}$ 
and a deceleration parameter {\it q}$_{0}$ = 0.},
the radio flux density at the wavelength of 6 cm, the ratio
of radio to optical flux density, the 60 $\mu$m luminosity, 
the [O {\sc iii}] luminosity, and the references for the 
[O {\sc iii}] luminosity and the emission-line flux ratios are given 
for each galaxy.
Those magnitudes are taken from V\'{e}ron-Cetty \& V\'{e}ron (1998), who
mentioned that they had chosen the magnitudes in the smallest possible 
diaphragm as they were interested in the nuclei rather than 
in the galaxy itself.
Table 3 describes the references for Table 5.

The emission-line flux ratios for each object are given in Table 6.
Each ratio is the averaged value among the references.
Since it is often difficult to measure the narrow Balmer component for S1s
accurately, there might be the systematic error if we make reddening
corrections using the Balmer decrement method (e.g., Osterbrock 1989) 
for all the types of Seyferts. Therefore we do not make the
reddening correction for the objects in our sample.
The effect of dust extinction on our result
is discussed in section 3.4.

\subsection{Selection Bias}

Because we do not impose any selection criteria upon our sample,
it is necessary to check whether or not the various samples are 
appropriate for our comparative study.
Systematic difference of the redshift distribution, the intrinsic
AGN power distribution, and the excitation degree of the narrow-line region
(NLR) gas among the various Seyfert types may cause possible biases, thus we 
investigate these distributions.

\subsubsection{Redshift}

The average redshifts and 1$\sigma$ deviations for each type are 
0.0351$\pm$0.0315 for the NLS1s, 
0.0550$\pm$0.0450 for the BLS1s,
0.0378$\pm$0.0401 for the S1.5s, 
0.0243$\pm$0.0315 for the S2$^+$s, and
0.0353$\pm$0.0309 for the S2$^-$s.
We show the histograms of the redshift in Figure 1.
It is noted that the average redshifts of the S1 and the S1.5 sample are
a little higher than those of the other samples.
In order to investigate whether or not the frequency distributions of
the redshift are statistically different among the types of Seyferts,
we apply the Kolmogorov-Smirnov (KS) statistical test (Press et al. 1988). 
The null hypothesis is that the redshift distributions
among the NLS1s, the BLS1s, the S1.5s,
the S2$^+$s and the S2$^-$s come from the same underlying population.
The results are summarized in Table 7. 
We give the KS probabilities for the class of S2$_{\rm total}$,
which means S2$^+$ and S2$^-$ since the numbers of 
the S2$^+$s and the S2$^-$s in our sample are not so large.
We give two KS probabilities for each combination; the first line 
gives the KS probabilities for the case including the radio-loud objects
while the second line gives those for the case without the radio-loud objects.
The results are nearly the same for these two cases in each combination.
The results of the KS test suggest that the redshifts of the S1s are
systematically higher than those of the other samples.

In this paper, our main attention is addressed to the visibility of
the torus HINER emission among the different Seyfert types.
Since the torus HINER is located in the inner 1 pc region around
the central engine, the larger average redshift of the S1s may not
affect the visibility of the torus HINER. If the S1s could have 
intense circumnuclear star-forming regions, such emission would
contribute to the line emission. However, since it is known that
S1s tend to have few such circumnuclear star-forming regions
(Pogge 1989; Oliva et al. 1995; Heckman et al. 1995;
Gonz\'{a}lez Delgado et al. 1997;  Hunt et al. 1997),
such contamination is expected to be negligibly small. Therefore,
we conclude that our later analyses are free from the redshift difference
among the samples.

\begin{figure*}
\epsscale{0.5}
\plotone{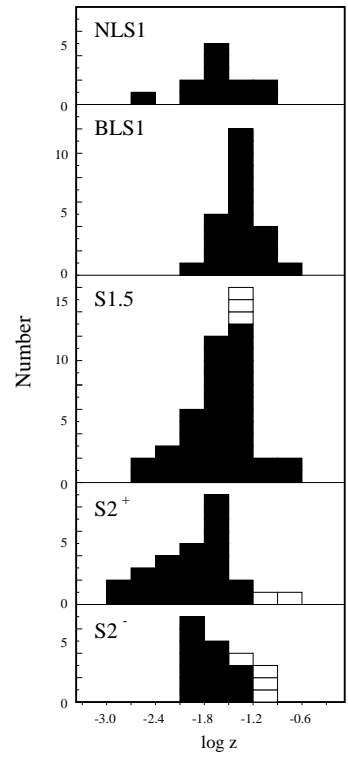}
\caption{
Frequency distributions of the redshift among each class of Seyfert nuclei.
The radio-loud galaxies are shown in white.
\label{fig1}}
\end{figure*}

\subsubsection{Luminosity}

\begin{figure*}
\epsscale{0.5}
\plotone{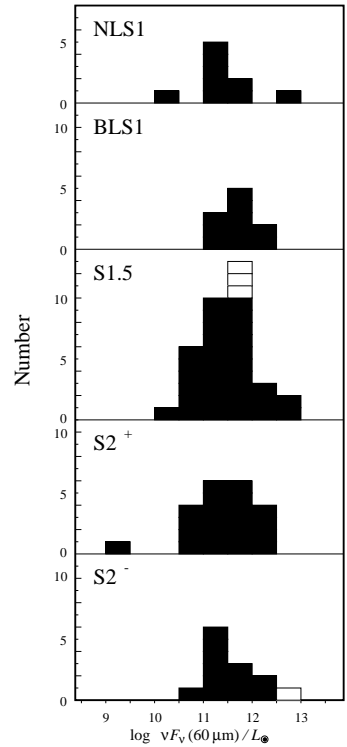}
\caption{
Frequency distributions of the 60$\mu$m luminosity among each class of 
Seyfert nuclei. The radio-loud galaxies are shown in white.
\label{fig2}}
\end{figure*}

\begin{figure*}
\epsscale{1.2}
\plotone{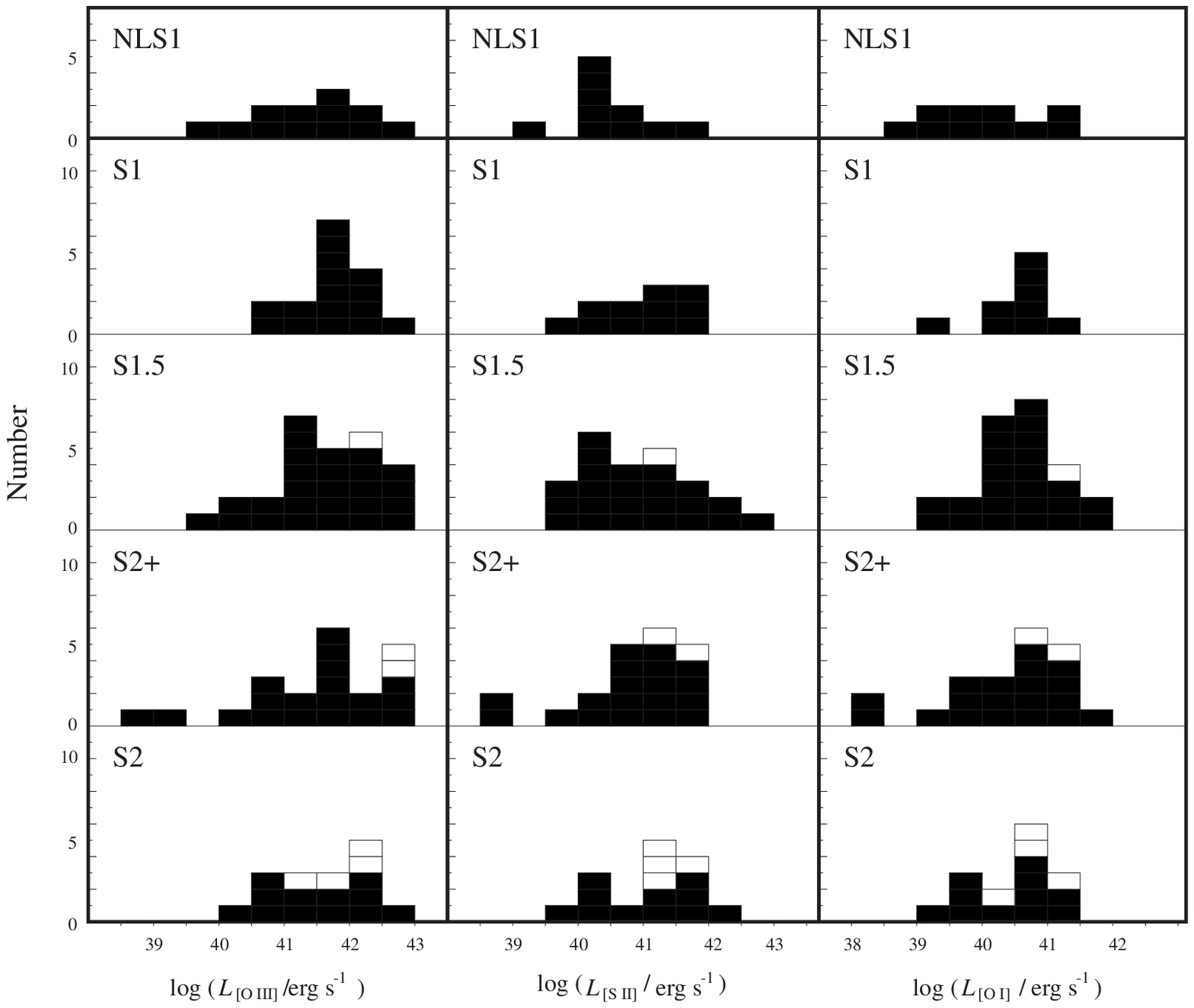}
\caption{
Frequency distributions of the luminosities of the low-ionization 
emission lines among each class of Seyfert nuclei. 
The radio-loud galaxies are shown in white.
a) The luminosity of [O {\sc iii}].
b) The luminosity of [S {\sc ii}].
c) The luminosity of [O {\sc i}].
\label{fig3}}
\end{figure*}

Current unified model of Seyfert galaxies require anisotropical nuclear 
radiation. This may cause systematical differences of intrinsic AGN power 
highly depending on selection criteria. Comparison of emission lines 
among different Seyfert types might suffer from this bias of intrinsic 
luminosity. Therefore, we investigate whether or not the intrinsic 
AGN power is systematically different among the different Seyfert types
using the luminosities which are regarded as isotropic emission reprocessed
from the nuclear radiation. We use {\it IRAS} 60$\mu$m and low-ionization
emission lines as such isotropic emission.

We firstly check the distributions of the 60$\mu$m luminosity 
among the samples.
The 60$\mu$m luminosity is thought to scale the nuclear continuum 
radiation which is 
absorbed and re-radiated by the dusty torus. Therefore the distribution
of the 60$\mu$m luminosity reflects that of the intrinsic luminosity.
The histograms of the 60$\mu$m luminosity are shown in Figure 2.
There appears to be no systematic difference among the types of Seyferts.
We apply the KS test where the null hypothesis is that the distribution of the
60$\mu$m luminosity among the various types of Seyferts come from the same
underlying population. The results suggest that there is no systematic
difference of the 60$\mu$m luminosity among the samples (see Table 8).
This means that there is no bias concerning to the intrinsic luminosity
in our sample.

However, the 60$\mu$m luminosity might be contaminated with the influence of
circumnuclear star formation. 
Hence we secondly investigate the luminosity of the 
low-ionization emission lines.
Because most of the flux of the low-ionization emission lines is radiated 
from the NLRs, it is thought to be almost independent 
with the viewing angle. Therefore the luminosity of a low-ionization 
emission line is a good tool 
to investigate the intrinsic power of the AGN.
As shown in Figure 3, the intensity distributions of the
low-ionization emission lines appear to be indistinguishable among the samples.
We apply the KS test where the null hypothesis is that the
luminosity of the low-ionization emission lines among various types of 
Seyferts come from the same underlying population.
The results suggest that there is no systematic
difference of the luminosity of the low-ionization emission lines 
among the samples (see Table 9).
Therefore we conclude that there is little 
difference of the distribution of the intrinsic luminosity among the samples.

\subsubsection{Excitation of the NLR gas}

\begin{figure*}
\epsscale{1.7}
\plotone{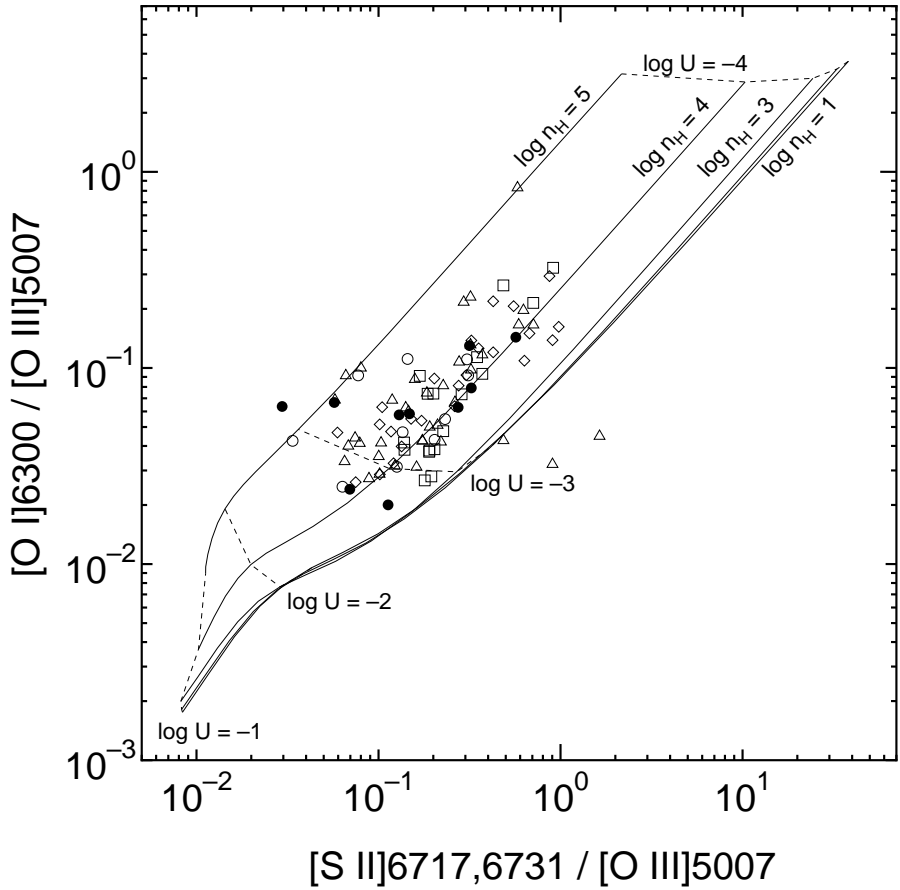}
\caption{
The photoionization model is drown in this diagram of 
[S {\sc ii}]/[O {\sc iii}] versus [O {\sc i}]/[O {\sc iii}].
The calculation are performed using the spectral synthesis model {\it CLOUDY}
version 90.04 (Ferland 1996). Here we assume that a uniform gas density,
solar abundance, dust-free gas with plane-parallel geometry is ionized by a 
`mean Seyfert' IR to gamma-ray continuum (with $\alpha_{\rm uv-x}$ = --1.4
in the EUV; see Ferland 1996). The free parameters for the calculations are
the hydrogen density of the cloud {\it n$_{\rm H}$} and the ionization 
parameter {\it U}, which is defined as the ratio of the ionizing photon 
density to the electron density. The values of there parameters are shown 
in the figure. The observed data of the NLS1s, the BLS1s, the S1.5s, 
the S2$^+$s, and the S2$^-$s are shown by filled circles, open ones,
triangles, diamonds, and squares, respectively.
\label{fig4}}
\end{figure*}

There is another problem concerning our comparisons in this paper.
In our study, we assume that the excitation degree of the NLRs is 
similar among the samples when we compare various line ratios.
In order to confirm the validity of this assumption,
we investigate whether or not the physical property
of the NLRs is different among the various types of Seyferts.
In Figure 4, we show the diagram of the intensity ratios of 
[S {\sc ii}]/[O {\sc iii}] versus [O {\sc i}]/[O {\sc iii}].
The diagram shows that there is little difference of the excitation 
degree of the NLRs among the types of Seyferts (see also Cohen 1983).
This guarantees the validity of the statistical comparisons in our study.

\section{RESULTS}

Emission-line ratios of AGNs 
have been often discussed in the form normalized
by the narrow component of Balmer lines; e.g., [O {\sc iii}]/H$\beta$,
[N {\sc ii}]/H$\alpha$, and so on (e.g., Veilleux \& Osterbrock 1987).
However, since we investigate emission-line properties
of S1s together with S2s,
we cannot use the usual emission-line ratios in our analysis.
Therefore, following the manner of MT98a, we investigate
intensity ratios between a HINER line and a low-ionization forbidden
emission line which is thought to be independent of the viewing-angle.

\subsection{The Relative Strength of the [Fe {\sc vii}] Emission}

\begin{figure*}
\epsscale{1.2}
\plotone{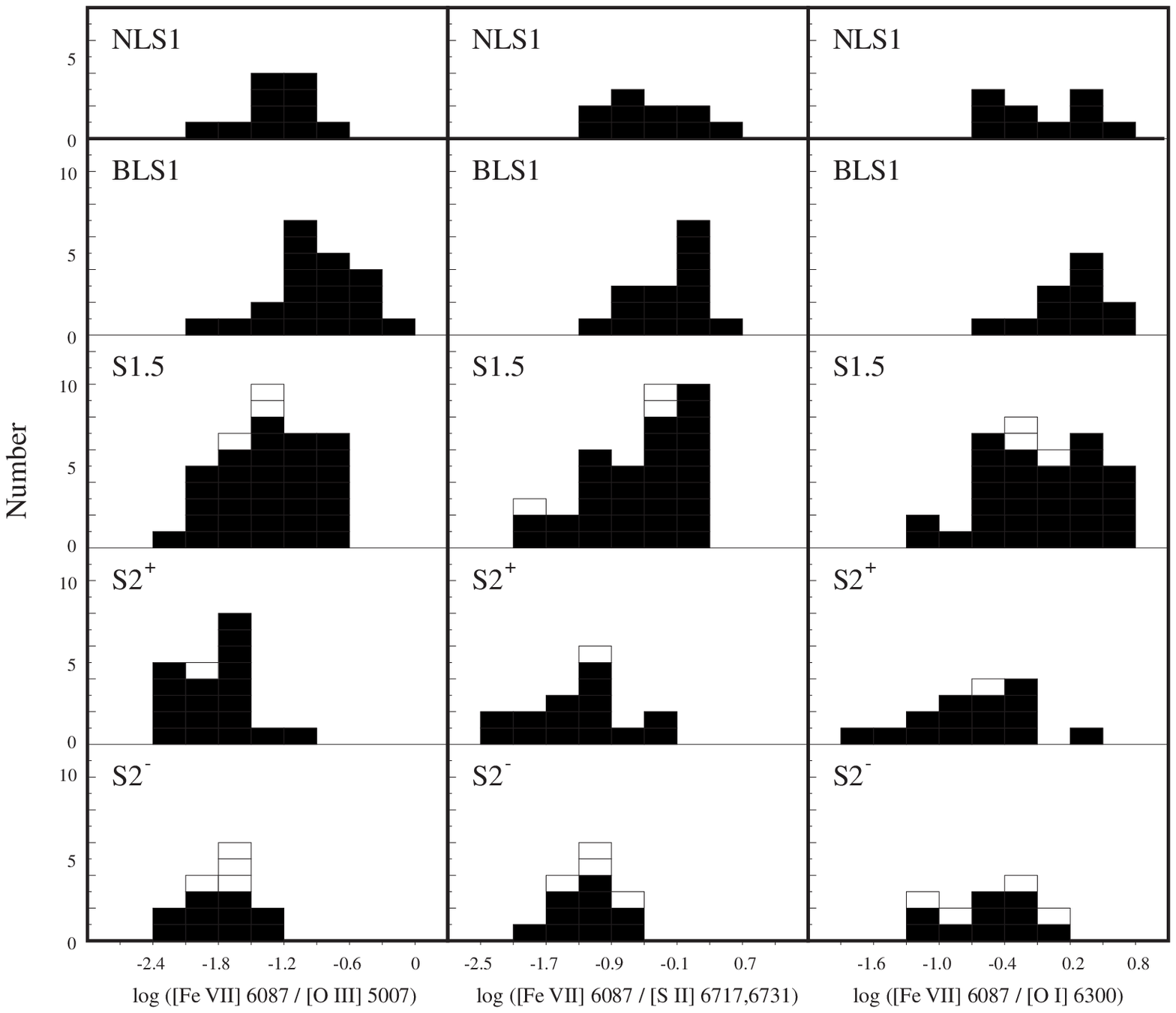}
\caption{
Frequency distributions of the line ratio of [Fe {\sc vii}]
to low-ionization emission lines among the NLS1s, the BLS1s, 
the S1.5s, the S2$^+$s, and the S2$^-$s. 
The number of radio-loud galaxies are shown in white.
Left) The line ratio of [Fe {\sc vii}]/[O {\sc iii}].
Middle) The line ratio of [Fe {\sc vii}]/[S {\sc ii}].
Right) The line ratio of 
[Fe {\sc vii}]/[O {\sc i}].
\label{fig5}}
\end{figure*}

\begin{figure*}
\epsscale{1.8}
\plotone{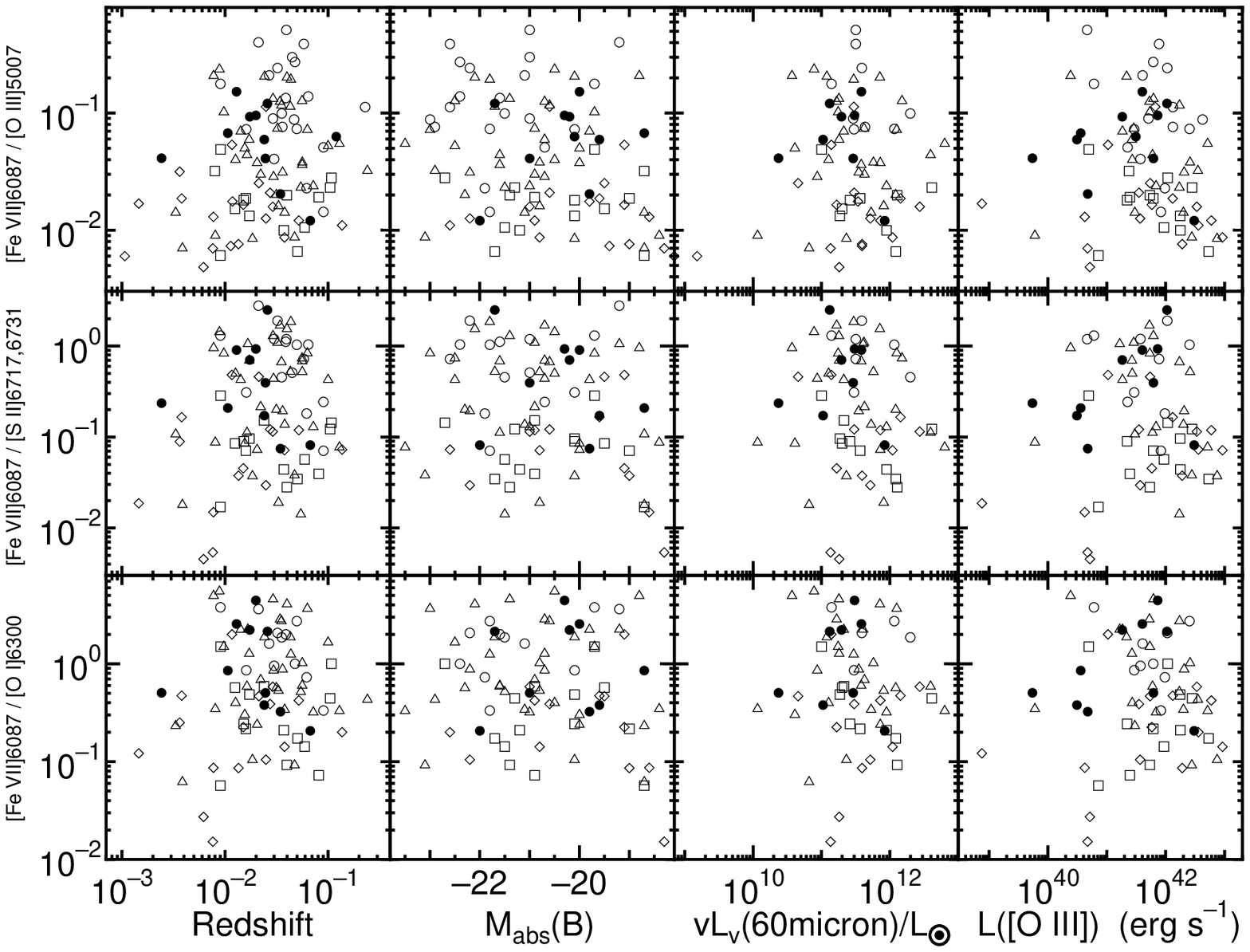}
\caption{
Diagrams of the redshift, the absolute {\it B} magnitude, the 60$\mu$m 
luminosity, and the [O {\sc iii}] luminosity versus the intensity ratio of
[Fe {\sc vii}] to low-ionization emission lines. The symbols are the same as 
in Figure 4. The ordinate of each diagram is the line ratio of
upper) [Fe {\sc vii}]/[O {\sc iii}],
middle) [Fe {\sc vii}]/[S {\sc ii}], and
lower) [Fe {\sc vii}]/[O {\sc i}].
\label{fig6}}
\end{figure*}

\begin{figure*}
\epsscale{1.2}
\plotone{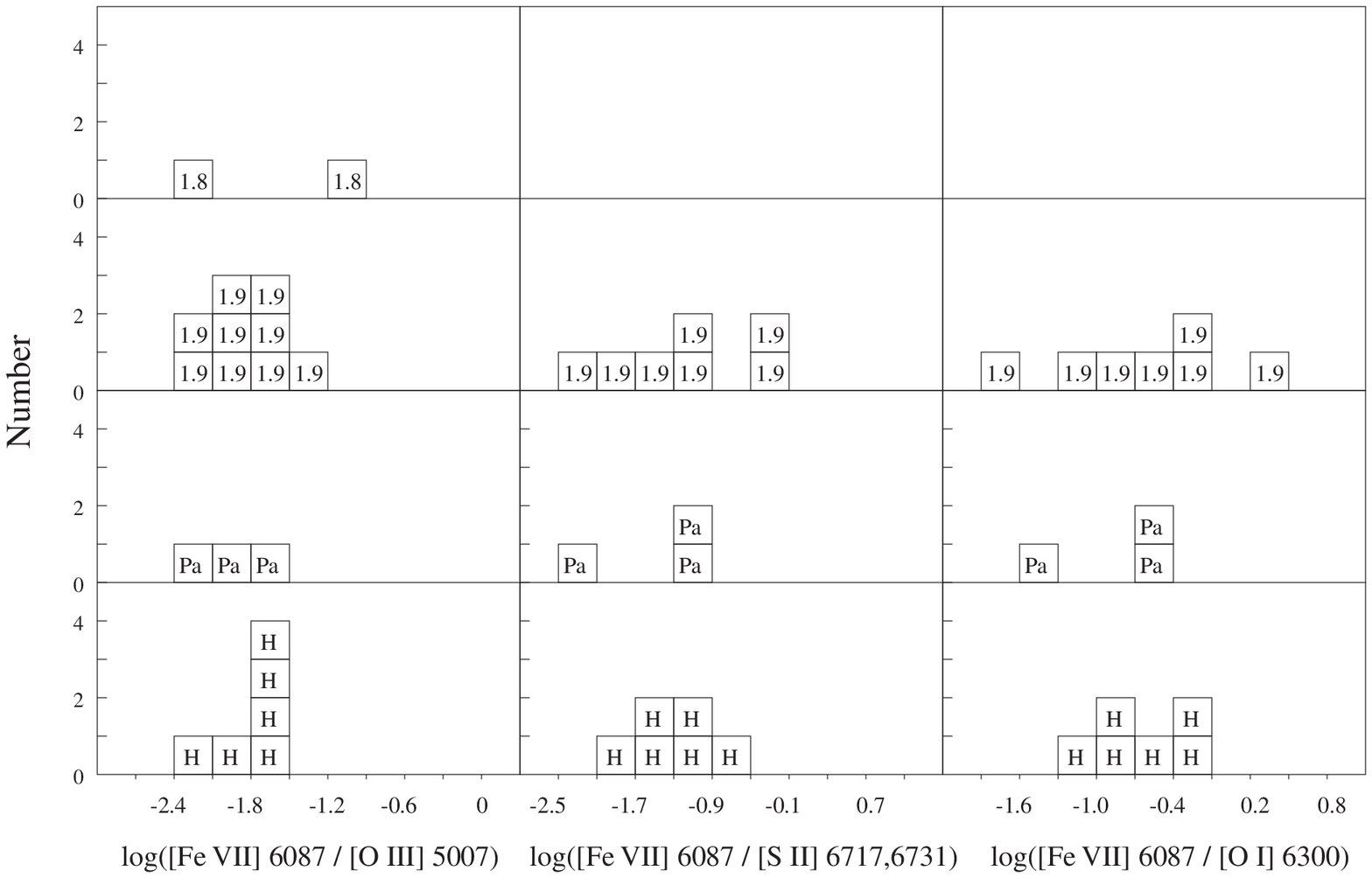}
\caption{
Frequency distributions of the line ratio of [Fe {\sc vii}] to
low-ionization emission lines concerning to the S2$^+$s.
``1.8'' means S1.8, ``1.9'' means S1.9, 
``Pa'' means S2$_{\rm NIR-NLR}$, and
``H'' means S2$_{\rm HBLR}$.
Left)   The line ratio of 
              [Fe {\sc vii}]/[O {\sc iii}].
Middle) The line ratio of 
              [Fe {\sc vii}]/[S {\sc ii}].
Right)  The line ratio of 
              [Fe {\sc vii}]/[O {\sc i}].
\label{fig7}}
\end{figure*}

We show the histograms of 
the line ratios of [Fe {\sc vii}] to [O {\sc iii}],
[S {\sc ii}] and [O {\sc i}],
for the NLS1s, the BLS1s, the S1.5s, the S2$^+$s, and the S2$^-$s in Figure 5.
Both the NLS1s and the BLS1s tend to have stronger [Fe {\sc vii}] emission
than the S2$^+$s and the S2$^-$s, being consistent with the result of MT98a.
It is interesting to note that the S1.5s show a marginal nature between
the S1s and the S2s. 

In order to investigate whether or not the differences
of the emission-line ratios among the samples are statistically real,
we apply the KS test. 
The null hypothesis is that the observed distributions
of the intensity ratios of [Fe {\sc vii}] to the low-ionization
emission lines among the NLS1s, the BLS1s, the S1.5s,
the S2$^+$s and the S2$^-$s come from the same underlying population.
The results are summarized in Table 10. 
It is noted that there is almost no difference between the KS probabilities
in the case of including the radio-loud galaxies and excluding those objects.

The KS test leads to the following results.
1) Both the NLS1s and the BLS1s have higher [Fe {\sc vii}] strengths
than the S2$^+$s and the S2$^-$s although the statistical significance is 
marginally low for the NLS1s when we use the [Fe {\sc vii}]/[O {\sc i}] ratio.
2) There is no statistical difference in the relative [Fe {\sc vii}] 
strength between the NLS1s and the BLS1s.
3) There is no statistical difference in the relative [Fe {\sc vii}] 
strength between the S2$^+$s and the S2$^-$s.
4) There is no statistical difference in the relative [Fe {\sc vii}] 
strength between the S1.5s and the S1s (i.e., the NLS1s and the BLS1s)
although the statistical significance is marginally low for the BLS1s
when we use the [Fe {\sc vii}]/[O {\sc iii}] ratio. And,
5) the S1.5s have higher [Fe {\sc vii}] strengths
than the S2$^+$s and the S2$^-$s. 
Because the frequency distributions of the luminosity of the low-ionization
emission lines are indistinguishable among the samples (see Figure 3),
the excess of the line ratios of [Fe {\sc vii}] to the low-ionization
emission lines in both the NLS1s and the BLS1s with respect to the S2$^+$s
and the S2$^-$s is thought to be due 
not to the depression of the emission of the low-ionization emission lines
but to the excess of the [Fe {\sc vii}] emission.

It is considered that the inclination effect is responsible for the above
results. In order to confirm this, we investigate
whether or not the relative
strength of the high-ionization emission lines correlate with the 
redshift or the intrinsic power of each AGN.
As shown in Figure 6, there appears to be no correlation between the relative 
[Fe {\sc vii}] strength and the redshift, the absolute {\it B} magnitude,
the 60$\mu$m luminosity, and the [O {\sc iii}] luminosity.
Therefore the various comparisons of the emission-line
flux ratios described in this paper are thought to be valid although 
there is slight difference in the average redshifts of the samples
(see Section 2.3.1).

We also investigate the HINER properties of the S1.8s, S1.9s,
S2$_{\rm NIR-BLR}$ and S2$_{\rm HBLR}$, respectively.
The results are shown in Figure 7.
There can be seen no systematic trend.
Therefore, these four sub-types are indistinguishable
in the HINER properties.

\subsection{The Relative Strength of the [Fe {\sc x}] Emission}

\begin{figure*}
\epsscale{1.2}
\plotone{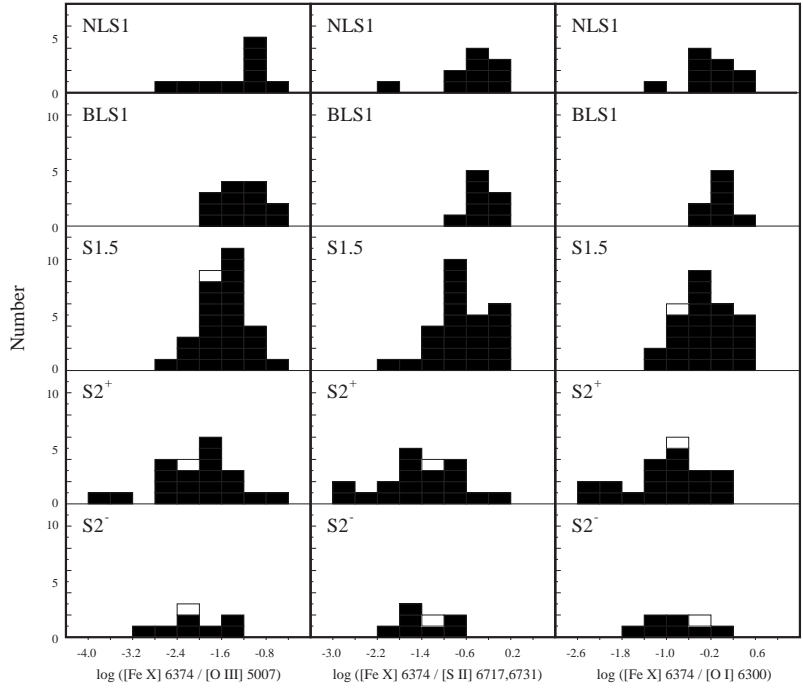}
\caption{
Frequency distributions of the line ratio of [Fe {\sc x}]
to low-ionization emission lines among each class of Seyfert nuclei.
The number of radio-loud galaxies are shown in white.
Left) The line ratio of [Fe {\sc x}]/[O {\sc iii}].
Middle) The line ratio of [Fe {\sc x}]/[S {\sc ii}].
Right) The line ratio of 
[Fe {\sc x}]/[S {\sc ii}].
\label{fig8}}
\end{figure*}

\begin{figure*}
\epsscale{1.8}
\plotone{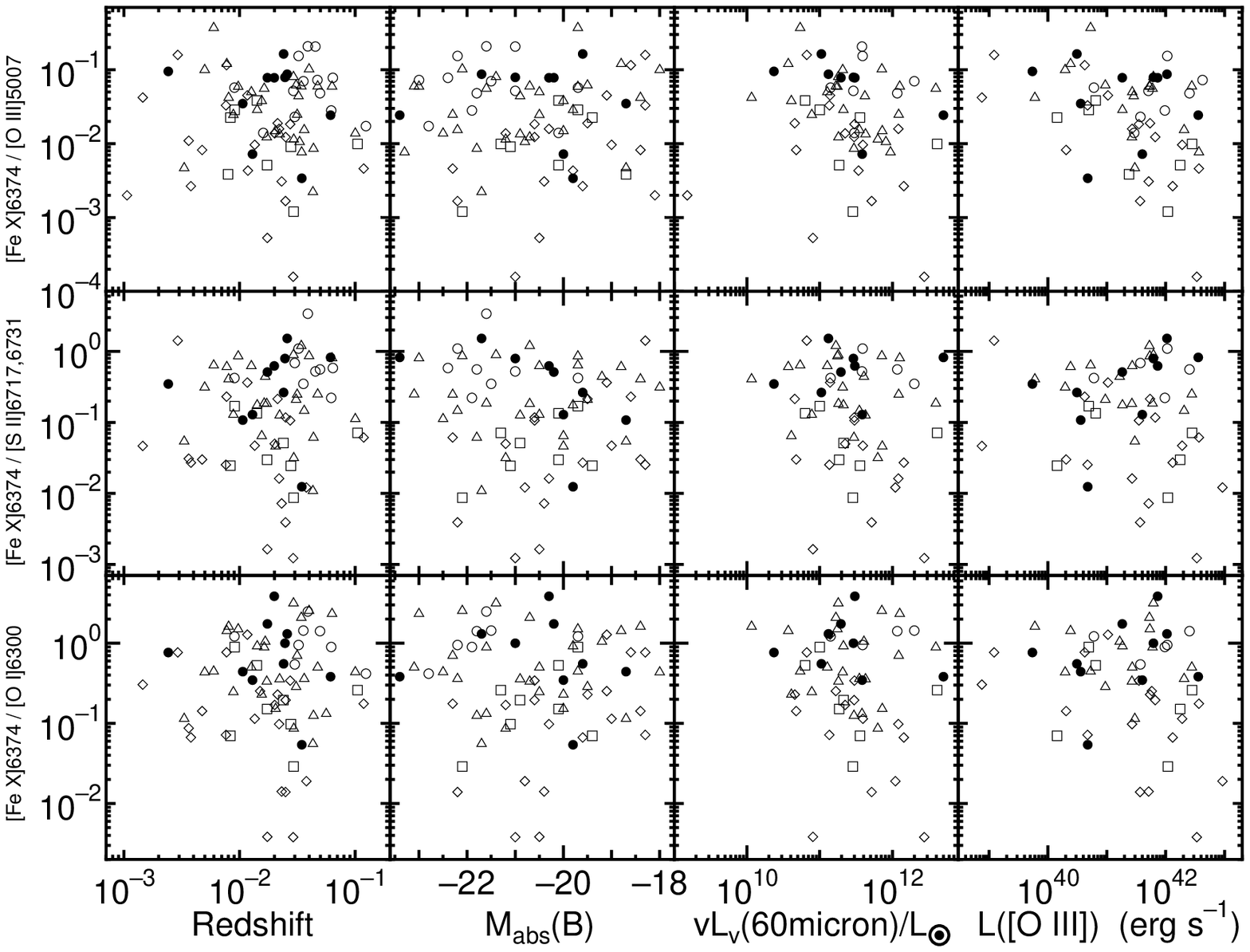}
\caption{
Diagrams of the redshift, the absolute {\it B} magnitude, the 60$\mu$m 
luminosity, and the [O {\sc iii}] luminosity versus the intensity ratio of
[Fe {\sc x}] to low-ionization emission lines. The symbols are the same as 
in Figure 4. The ordinate of each diagram is the line ratio of
upper) [Fe {\sc x}]/[O {\sc iii}],
middle) [Fe {\sc x}]/[S {\sc ii}], and
lower) [Fe {\sc x}]/[O {\sc i}].
\label{fig9}}
\end{figure*}

\begin{figure*}
\epsscale{1.2}
\plotone{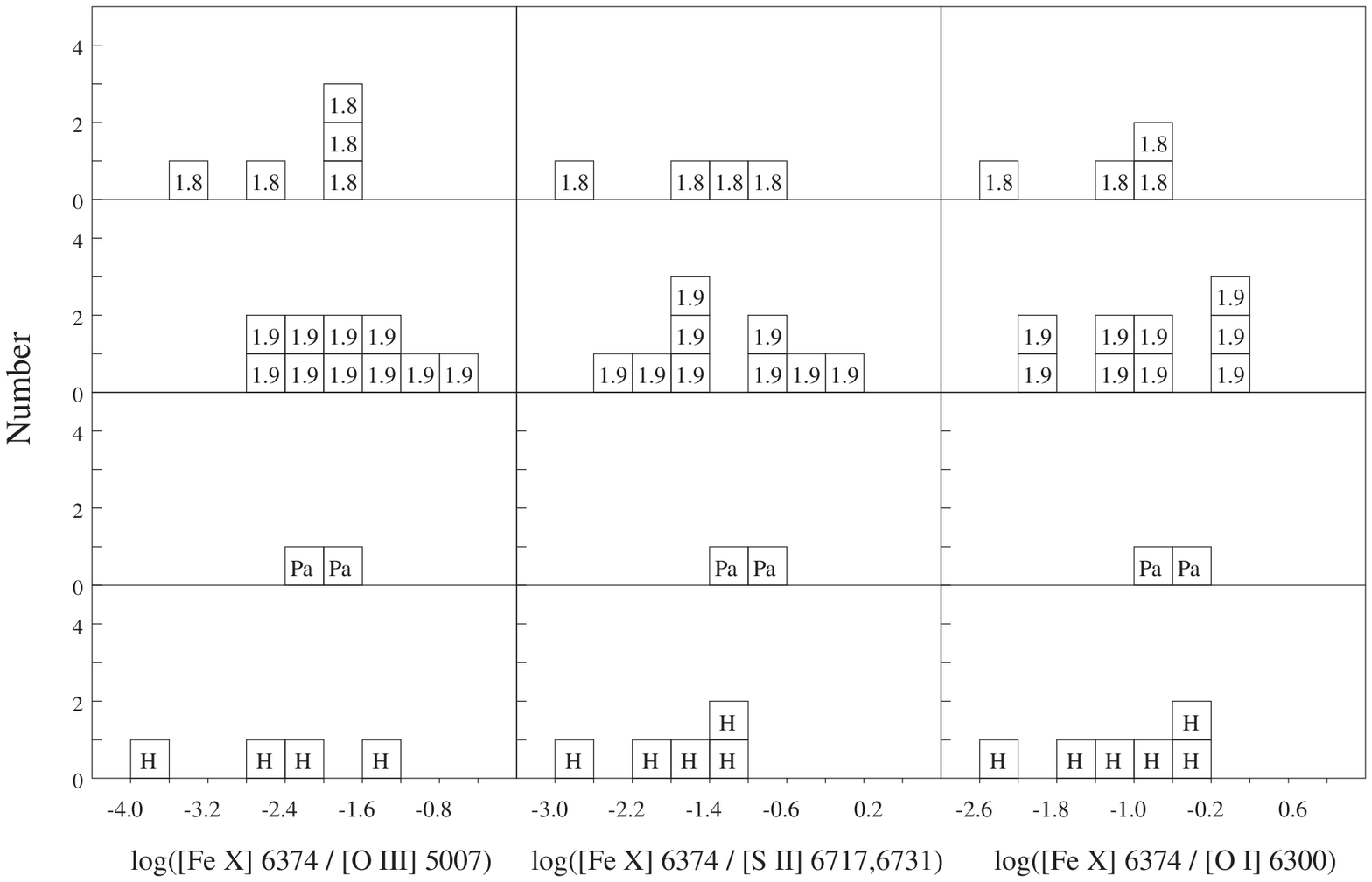}
\caption{
Frequency distributions of the line ratio of [Fe {\sc x}] to
low-ionization emission lines concerning to the S2$^+$s.
The notations are same in Figure 7.
Left)   The line ratio of 
              [Fe {\sc x}]/[O {\sc iii}].
Middle) The line ratio of 
              [Fe {\sc x}]/[S {\sc ii}].
Right)  The line ratio of 
              [Fe {\sc x}]/[O {\sc i}].
\label{fig10}}
\end{figure*}

We present the histograms of the intensity ratios of [Fe {\sc x}]
to low-ionization emission lines in Figure 8.
We apply the KS test where the null hypothesis is that
the observed distributions
of the intensity ratios of [Fe {\sc x}] to the low-ionization
emission lines among various types of Seyferts come from the same
underlying population. The results are given in Table 11.
There is also no difference between the KS probabilities in the case of 
including the radio-loud galaxies and excluding those objects.

The KS test leads to the following results.
1) Both the NLS1s and the BLS1s have higher [Fe {\sc x}] strengths
than the S2$^+$s and the S2$^-$s. However, the statistical significance
is much worse than that using the [Fe {\sc vii}] emission.
2) There is no statistical difference in the relative [Fe {\sc x}] 
strength  between the NLS1s and the BLS1s.
3) There is no statistical difference in the relative [Fe {\sc x}] 
strength between the S2$^+$s and the S2$^-$s
4) There is no statistical difference in the relative [Fe {\sc x}] 
strength between the S1.5s and the S1s (i.e., the NLS1s and the BLS1s). And,
5) it is not clear whether or not there is statistical difference 
in the relative [Fe {\sc x}] 
strength between the S1.5s and the S2s (i.e., S2$^+$s + S2$^-$s).
These results are not consistent with those using the [Fe {\sc vii}] emission. 
This point will be discussed in next section.

In Figure 9, we show the diagrams of the redshift, the absolute {\it B} 
magnitude, the 60$\mu$m luminosity, and the [O {\sc iii}] luminosity
versus the relative [Fe {\sc x}] strength.
Similar to the case mentioned in the previous section, there is no correlation
between the relative [Fe {\sc x}] strength and the redshift, the 
absolute {\it B} 
magnitude, the 60$\mu$m luminosity, and the [O {\sc iii}] luminosity.
This result also assures the validity of our comparative study.

We also investigate the HINER properties of the S1.8s, S1.9s,
S2$_{\rm NIR-BLR}$ and S2$_{\rm HBLR}$, respectively.
The result is shown in Figure 10.
Again, there can be seen no systematic trend in this figure.
Therefore, these four subclasses are indistinguishable
in the HINER properties.

\subsection{[Fe {\sc vii}] versus [Fe {\sc x}]}

\begin{figure*}
\epsscale{0.5}
\plotone{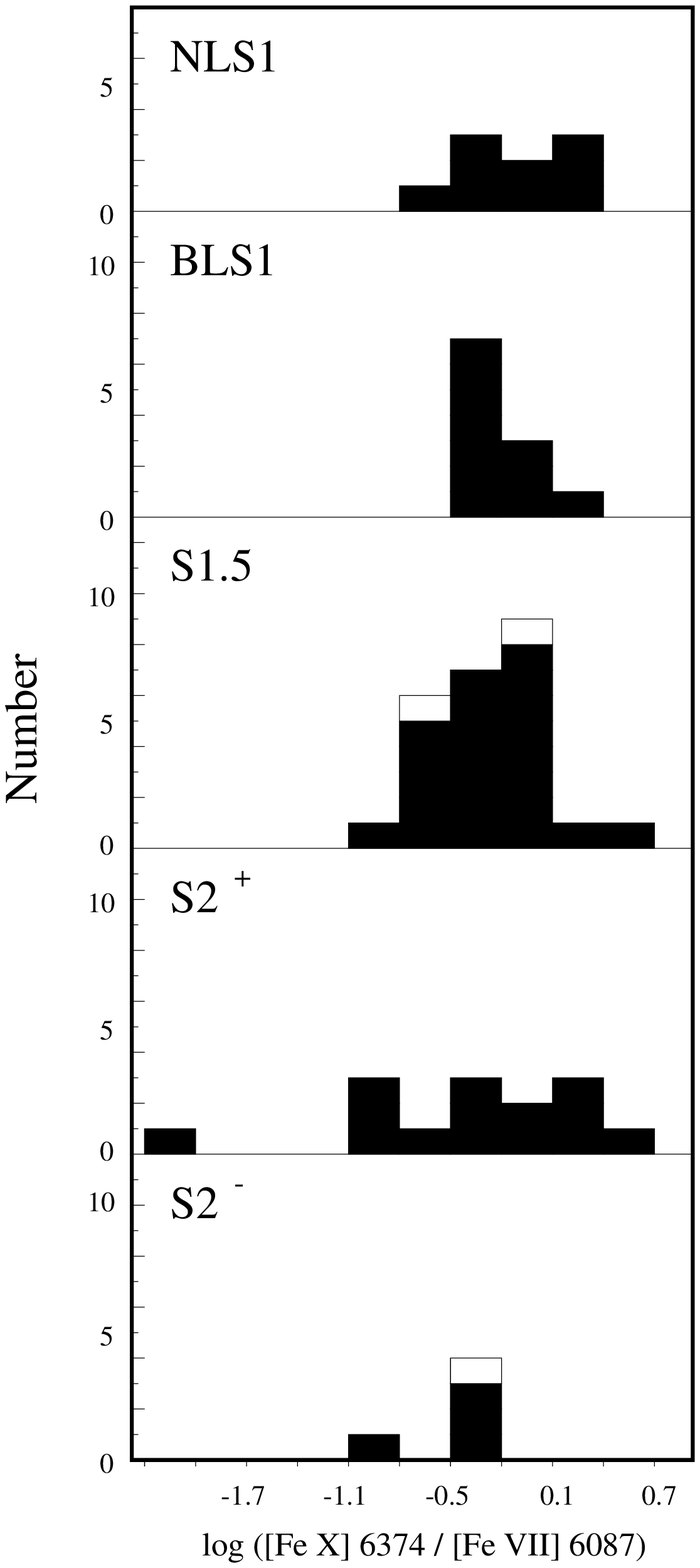}
\caption{
Frequency distributions of the line ratio of [Fe {\sc xi}]
to [Fe {\sc x}] among each class of Seyfert nuclei.
The number of radio-loud galaxies are shown in white.
\label{fig11}}
\end{figure*}

We investigate whether or not the [Fe {\sc x}]/[Fe {\sc vii}] ratio
is different among the samples. The frequency distributions
of this ratio are shown in Figure 11.
We apply the KS test where the null hypothesis is that
the observed distributions
of the [Fe {\sc x}]/[Fe {\sc vii}] ratio
among the various types of Seyferts come from the same
underlying population. The results are given in Table 12.
Although there seem to be a marginal tendency that the NLS1s
have higher [Fe {\sc x}]/[Fe {\sc vii}] ratios than the
other types of Seyferts, the KS test shows that
this is not statistically real.

\subsection{Effects of the Dust Extinction}

As mentioned in section 2.2, no reddening correction has 
been made for all the observed emission line ratios analyzed here.
However, it is known that the dust extinction is larger
on average in S2s than in S1s (Dahari \& De Robertis 1988a, 1988b).
In order to see how the extinction affects the line ratios,
we summarize the shifts of the line ratios for the following
three cases; $A_V$ = 1.0, 5.2, and 10.0 (see Table 13).
The case of $A_V$ = 5.2 corresponds to that of the Circinus galaxy
(Oliva et al. 1994).
In these estimates, we use the Cardelli's extinction curve
(Cardelli, Clayton, \& Mathis 1989).

Since the wavelength of [O {\sc i}] is relatively close to those
of [Fe {\sc vii}] and [Fe {\sc x}], the effect of dust extinction
is negligibly small even for the case of  $A_V$ = 10.0
when the [O {\sc i}] intensity is used as a normalizer.
When normalized by [O {\sc iii}],
the effect of dust extinction leads to higher 
[Fe {\sc vii}]/[O {\sc iii}] and [Fe {\sc x}]/[O {\sc iii}] ratios.
Because the dust extinction is larger on average in S2s than in S1s,
the observational results show that these ratios are higher in the S1s 
than in the S2s even if the effect of the dust extinction is 
taken into account.
On the other hand, the excess of [Fe {\sc vii}]/[S {\sc ii}] 
in both the NLS1s and the BLS1s with respect to the S2s would be extinguished
if the extinction of the S2s is systematically larger (e.g.,
$A_V$ = 10) than that of the S1s.
However, the average difference of the extinction between S1s and S2s
is about 1 mag (Dahari \& De Robertis 1988a; see
also De Zotti \& Gaskell 1985). Hence it is unlikely that
the S2s analyzed here suffer from such larger extinction systematically.
Therefore, we conclude that the results obtained in our analysis 
are not so seriously affected by the dust extinction.

\section{DISCUSSION}

\subsection{The HINER in the NLS1s}

\begin{figure*}
\epsscale{1.0}
\plotone{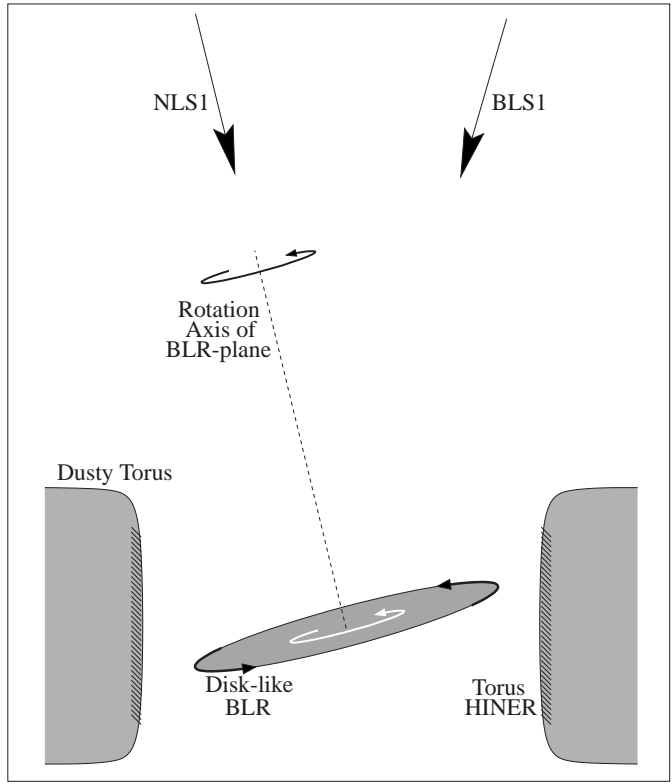}
\caption{
A schematic illustration of the difference of the viewing angle
between BLS1s and NLS1s.
\label{fig12}}
\end{figure*}

Our analysis has shown that;
1) the NLS1s have higher [Fe {\sc vii}] and [Fe {\sc x}] strengths
than the S2$^+$s and the S2$^-$s although the statistical significance
is worse when using the [Fe {\sc x}] emission, and 
2) there are no statistical differences in the relative strength of
[Fe {\sc vii}] and [Fe {\sc x}]
between the NLS1s and the BLS1s.
Several previous works suggested that strong high-ionization 
emission lines are often seen in NLS1s (Davidson \& Kinman 1978;
Osterbrock \& Pogge 1985; Nagao et al. 2000). Our analysis has
statistically confirmed for the first time that the HINER emission lines 
of the NLS1s are significantly stronger than those of the S2s.
Accordingly this suggests that the NLS1s are viewed from a more face-on
view toward dusty tori than the S2s.
On the other hand, the second result means that 
there is no systematic difference in the viewing angle toward 
the dusty torus 
between the NLS1s and the BLS1s from a statistical point of view.

Many theoretical models have been proposed to explain the properties of NLS1s
(e.g., Boller et al. 1996; Taniguchi et al. 1999 and references therein).
Any model is required to satisfy the statistical properties of the HINER
presented in this paper; i.e., 
the narrow line width of NLS1s cannot be explained assuming 
the obscuration of broad component with dusty torus.
For example, Giannuzzo \& Stripe (1996) mentioned a possibility 
that the NLS1s may be objects seen from relatively large inclination 
angles and thus only outer parts of the BLR can be seen,
being responsible for the narrow line width.
However such models appear difficult to explain
the property of HINER in the NLS1s consistently.

Since our second result means that the viewing angle toward the dusty 
torus is nearly the same on average between the NLS1s and the BLS1s,
it is possible to propose the following model if the BLR observed in
optical spectra has a disk-like configuration. 
Suppose that the rotational axis 
of the BLR is different from that of the dusty torus.
In this case, the BLR line width of S1s depends on the viewing angle 
toward the BLR disk. However, the line width does not depend on the
the viewing angle toward the dusty torus unless the BLR is not hidden
by the dusty torus. This model is schematically shown in Figure 12.
Our results appear consistent with this model.

It is noted that the BLR emission may arise from 
outer parts of a warped accretion disk
(Shields 1977; Nishiura, Murayama, \& Taniguchi 1998; 
see also for a review Osterbrock 1989).
Such warping of accreting gas disks may be driven by the effect of
radiation pressure force (Pringle 1996, 1997).
Indeed, it has been recently shown that accreting gas clouds probed 
water vapor maser emission at 22 GHz show evidence for significant warping
(Miyoshi et al. 1995; Begelman \& Bland-Hawthorn 1997).
Therefore, it is likely that the rotation axis of the BLR is not necessarily
to align to that of the dusty torus.

Recently, Turner, George, \& Netzer (1999b) reported on the observation of
the NLS1 Akn 564. They estimated the viewing angle toward the accretion
disk $\sim$60\arcdeg \ using a model for asymmetric Fe K$\alpha$ line profile
and mentioned that this result is contrary to the hypothesis that NLS1s are
viewed from pole-on view.
However, if the rotation axis of the accretion disk is different from
those of the BLR and the dusty torus 
(e.g., Nishiura, Murayama, \& Taniguchi 1998),
their observation is not inconsistent with the HINER properties
presented in this paper.

\subsection{The HINER in the S1.5s}

The S1.5 is widely recognized as a distinct class of Seyferts
observationally
(Osterbrock \& Koski 1976; Cohen 1983). 
However, the nature of this type of Seyferts have not yet been
fully understood. Comparing the HINER properties of the S1.5s with
those of the other types of Seyferts, we discuss the nature of S1.5s.

Our analysis shows that;
1) there is no statistical difference in the relative [Fe {\sc vii}]
and [Fe {\sc x}] strengths between the S1.5s and the S1s
(i.e., NLS1s + BLS1s), but 
2) the S1.5s have higher [Fe {\sc vii}] strengths
than the S2$^+$s and the S2$^-$s although this tendency is not confirmed
in the relative [Fe {\sc x}] strength.
In summary, as shown in Figures 5 and 8, although 
the S1.5s have an intermediate
property in the HINER line strengths between the S1s and S2s,
the relative HINER line strengths cover the whole observed ranges
of both the S1s and the S2s.
Therefore, there are three alternative ideas to explain these
observational properties. The first idea is that
the S1.5s are seen from an
intermediate viewing angle between S1s and S2s; i.e.,
a significant part of the BLR is obscured by a dusty torus,
resulting in a composite profile consisting of both 
the narrow-line region (NLR) and BLR emission.
The second idea is that some S1.5s are basically S1s but
a significant part of the BLR emission is accidentally  obscured 
by dense, clumpy gas clouds.
The third idea is that some S1.5s are basically
S2s but a part of the BLR emission can be seen from 
some optically-thin regions of the dusty torus.

As shown in Figure 5, the majority of S1.5s has nearly the same
relative [Fe {\sc vii}] strengths as those of the S1s, being consistent
with the second idea. The remaining minority can be explained
either by the first idea or by the third one.
Yet, it seems important to mention that the origin of S1.5s may be
heterogeneous. Finally, it is also important to mention that the latter 
two ideas may explain why some Seyfert nuclei show the so-called type 
switching between S1 and S2; e.g., NGC 4151 (Penston \& Perez 1984; 
Ayani \& Maehara 1991).
The reason for this is as follows.
It seems likely that the dusty torus consists of rather small blobs which are 
orbiting around the central engine. When a blob is passing the line of sight
to the central engine, the BLR can be obscured if the blob is optically thick
enough to hide it. If we assume that the blob is located at a radial distance 
of 0.1 pc from the central engine and the mass of the supermassive black hole
is 10$^7$ M$_{\sun}$, the Keplerian velocity is estimated to be 
{\it V}$_{\rm rot} \sim$ 660 km s$^{-1}$. Since the typical time scale of the
observed type switchings is $\sim$ 10 years, this blob could move 2 $\times$ 
10$^{16}$ cm. This is almost consistent with the typical size of the BLR,
$\sim$ 0.01 pc (e.g., Peterson 1993). This idea also suggests that
a typical size of such blobs is $\sim$ 0.01 pc.

\subsection{The HINER in the ``S2$^+$s''}

As presented in section 3, there is no statistical difference
in the relative strengths of [Fe {\sc vii}] and [Fe {\sc x}]
between S2$^+$s and S2$^-$s .
Among the subtypes of S2$^+$s (i.e., S1.8, S1.9, S2$_{\rm NIR-BLR}$
and S2$_{\rm HBLR}$), there is no systematic difference
in the HINER properties (see Figures 7 and 10).  
On the other hand, the S2$^+$s have weaker [Fe {\sc vii}] strength
than the S1.5s (Table 10).
These facts suggest that there is a systematic difference in the viewing 
angle toward dusty tori between the S1.5s and the S2$^+$s; i.e.,
the S2$^+$s might be those which are seen with
large inclination angle and the emission radiated from the BLR reach
us through the occasionally thin dusty tori.
All these arguments imply that S2$^+$s are viewed from
large inclination angles, leading to more significant extinction
of the BLR emission with respect to BLS1s and S1.5s.
This appears consistent with earlier implications (e.g.,
Miller \& Goodrich 1990; Heisler, Lumsden, \& Bailey 1997).

\subsection{The Nature of HINER Traced by [Fe {\sc x}]}

As presented in sections 3.1 and 3.2,
both the NLS1s and the BLS1s have higher HINER emission-line
strengths than the S2$^+$s and the S2$^-$s. 
However, comparing the KS probabilities given in Table 10 and Table 11,
in particular those concerning to the relative intensities normalized
by [O {\sc iii}], this tendency is much more prominent 
in the analysis using [Fe {\sc vii}] rather than [Fe {\sc x}].
As proposed by MT98a, the excess [Fe {\sc vii}] emission in the S1s
appears attributed to the significant contribution from the torus 
HINER. Therefore, the weaker excess emission in [Fe {\sc x}]
implies that the major [Fe {\sc x}] emitting region
may be not the torus HINER but either the clumpy NLR HINER or
the extended HINER or both because the latter two HINERs show
less viewing angle dependence (see MT98a).

Another interesting property related to the [Fe {\sc x}] emission 
is the observed [Fe {\sc x}]/[Fe {\sc vii}] ratios (see Figure 11).
The average ratios are compared among the sample in Table 14.
It is remarkable that some S1s and S2s have very 
higher ratios; e.g., [Fe {\sc x}]/[Fe {\sc vii}] $\geq$ 1.
Therefore, in order to investigate the origin of the [Fe {\sc x}]
emission, it is interesting to compare the observed line ratios of
[Fe {\sc x}]/[Fe {\sc vii}] with several theoretical predictions.
Because simple one-zone models are known to predict too weak
high-ionization emission lines (e.g., Pelat, Alloin, \& Bica 1987;
Dopita et al. 1997), we investigate multi-component photoionization models; 
1) optically thin multi-cloud model (Ferland \& Osterbrock 1986), and
2) the locally optimally emitting cloud model (LOC model; 
Ferguson et al. 1997).
As shown in Table 14, these models predict smaller line ratios.
On the other hand, the low-density interstellar matter (ISM) model by 
Korista \& Ferland (1989) predicts higher line ratios of 
[Fe {\sc x}]/[Fe {\sc vii}].
They mentioned that such HINER will be observed out to 1 -- 2 kpc.
Indeed, extended HINERs have been found in some Seyfert galaxies;
NGC 3516 (Golev et al. 1995), Tololo 0109$-$383 (Murayama et al. 1998),
and NGC 4051 (Nagao et al. 2000).
Alternatively, shock models may also be responsible for the observed
higher [Fe {\sc x}]/[Fe {\sc vii}] ratios.
Viegas-Aldrovandi \& Contini (1989)
calculated those ratios introducing the shock component.
As shown in Table 14, such models can also explain those higher line ratios.

Which is the appropriate model for the higher line ratios of 
[Fe {\sc x}]/[Fe {\sc vii}], highly ionized low-density ISM or
shock-driven ionization?
In order to distinguish these two models, we compare the observed 
[Fe {\sc xi}]/[Fe {\sc x}] ratios with model results in Table 14.
The [Fe {\sc xi}] emission is observed in only 19 Seyfert nuclei
(e.g., Grandi 1978; Cohen 1983; 
Penston et al. 1984; Erkens et al. 1997).As shown in Table 14,
the low-density ISM models of Korista \& Ferland (1989) appear 
consistent with the observed [Fe {\sc xi}]/[Fe {\sc x}]ratios.
Though Viegas-Aldrovandi \& Contini (1989) did not calculate this line ratio,
Evans et al. (1999) mentioned that the ionization state 
in the emission-line region ionized by shocks is somewhat lower than
that ionized by a typical nonthermal continuum.
Therefore, it is likely that some Seyfert galaxies have the
extended HINER described by Korista \& Ferland (1989), 
being responsible for the unusually strong 
[Fe {\sc x}] emission. This idea also explains why the excess
[Fe {\sc x}] emission is less significant in the S1s 
than the excess [Fe {\sc vii}] emission.

\section{CONCLUDING REMARKS}

The anisotropic property of the radiation from the HINER traced 
[Fe {\sc vii}] reported by MT98a
has been statistically confirmed using the larger sample.
The line ratios of [Fe {\sc x}] to the low-ionization
emission lines show a rather isotropic property with respect to
those of [Fe {\sc vii}] to the low-ionization emission lines.
This may be interpreted by an idea that a significant fraction
of the [Fe {\sc x}] emission arises from low-density ISM
as suggested by Korista \& Ferland (1989). We note that 
the [Fe {\sc x}] emission is not suitable 
to investigate the viewing angle toward the dusty tori of Seyfert nuclei.

We have also investigated the HINER properties of the intermediate-type
of Seyfert nuclei.
Using the frequency distributions of the line ratios of 
[Fe {\sc vii}] to the low-ionization emission lines,
we find the following suggestions.
(1) The NLS1s are viewed from a more face-on orientation toward dusty tori
than the S2s.
(2) The line ratios of S1.5s are distributed in a wide range from the smallest
value of the S2s to the largest value of the S1s.
This suggests that the S1.5s are heterogeneous populations.
(3) The HINER properties of the S1.8s, the S1.9s and the objects showing 
a broad Pa$\beta$ line or polarized broad Balmer lines are considerably
different from those of the S1s.
These facts mean that the ``S2$^+$'' objects are those which are seen 
from a large inclination angle and their BLR emission comes through
optically-thin line of sights toward the dusty tori.

\acknowledgments

We would like to thank the anonymous referee for useful comments and 
suggestions and Yuji Ikeda, Shingo Nishiura and Yasuhiro Shioya 
for useful advice.
This research has made use of the NED (NASA extragalactic database) 
which is operated by the Jet Propulsion Laboratory, 
California Institute of Technology, under construct with the National
Aeronautics and Space Administration.
TM is supported by a Research Fellowship from the Japan Society for 
the Promotion of Science for Young Scientists.
A part of this work was made when YT visited the Astronomical Data 
Analysis Center (ADAC) of the National Astronomical Observatory
of Japan. YT thanks the staff of ADAC, in particular Shin-ichi Ichikawa,
for their kind hospitality.
This work was financially supported in part by Grant-in-Aids for the Scientific
Research (Nos. 10044052, and 10304013) of the Japanese Ministry of
Education, Culture, Sports, and Science.

\clearpage

\clearpage
%%%%%%%%%% TABLE 1 %%%%%%%%%%
\begin{deluxetable}{llrcl}
\tablenum{1}
\tablecaption{Our Classification Scheme \label{tbl-1}}
\tablewidth{0pt}
\tablehead{
\colhead{Class} &
\colhead{Property} &
\colhead{} &
\colhead{Our Notations} &
\colhead{}
}
\startdata
NLS1 & FWHM(H$\beta$) $<$ 2000 km s$^{-1}$                           & & NLS1 & \nl
     & [O {\sc iii}]$\lambda$5007/H$\beta$ $<$ 3                    & &      & \nl
     & strong Fe {\sc ii} emission                                  & &      & \nl
S1   & showing broad Balmer lines                                   & & BLS1 & \nl
S1.2 & intermediate between S1 and S1.5                             & & BLS1 & \nl
S1.5 & apparent narrow H$\beta$ profile superinposed on broad wings & & S1.5 & \nl
S1.8              & intermediate between S1.5 and S2                                     & S2$_{\rm RBLR}$ & S2$^{+}$ &S2$_{\rm total}$ \nl
S1.9              & broad component visible in H$\alpha$ but not in H$\beta$             & S2$_{\rm RBLR}$ & S2$^{+}$ &S2$_{\rm total}$ \nl
S2$_{\rm NIR-BLR}$& broad component visible in Pa$\beta$ but not in optical Balmer lines & S2$_{\rm RBLR}$ & S2$^{+}$ &S2$_{\rm total}$ \nl
S2$_{\rm HBLR}$   & broad component visible only in polarized Balmer lines               &                 & S2$^{+}$ &S2$_{\rm total}$ \nl
S2                & broad component invisible with any method                            &                 & S2$^{-}$ &S2$_{\rm total}$ \\
\enddata
\end{deluxetable}
%%%%%%%%%% TABLE 1 %%%%%%%%%%

%%%%%%%%%% TABLE 2 %%%%%%%%%%
\begin{deluxetable}{llcccccc}
\tablenum{2}
\tablecaption{Classification of Our Sample \label{tbl-2}}
\tablehead{
\colhead{Name} &
\colhead{Another Name} &
\colhead{DDR88\tablenotemark{a}} & 
\colhead{S89\tablenotemark{b}} & 
\colhead{W92\tablenotemark{c}} &
\colhead{CG94\tablenotemark{d}} &
\colhead{VCV98\tablenotemark{e}}  & 
\colhead{This paper} 
}
\startdata
\cutinhead{Radio Quiet AGN}
NGC 424	&Tololo 0109-383&\nodata	&\nodata&S2	&\nodata&S1.9	&S1.9 \nl
NGC 1019&\nodata	&\nodata	&\nodata&S1	&S1	&S1.5	&S1.5 \nl
NGC 1068& M 77		&S2		&\nodata&S2	&\nodata&S2$_{\rm HBLR}$	&S2$_{\rm HBLR}$ \nl
NGC 1566&\nodata	&S1		&\nodata&S1.5	&\nodata&S1.5	&S1.5 \nl
NGC 2110&\nodata	&S2		&\nodata&S2	&S2	&S1.9	&S1.9 \nl
NGC 2992&\nodata	&S2\tablenotemark{f}&\nodata&S1.9&\nodata&S1.9	&S1.9 \nl
NGC 3081&\nodata	&S2\tablenotemark{f}&\nodata&S2	&S2	&S2	&S2 \nl
NGC 3227&\nodata	&S1.5		&\nodata&S1.2	&\nodata&S1.5	&S1.5 \nl
NGC 3362&\nodata	&\nodata	&\nodata&\nodata&\nodata&S2	&S2 \nl
NGC 3393&\nodata	&\nodata	&\nodata&\nodata&\nodata&S2	&S2 \nl
NGC 3516&\nodata	&S1.5		&\nodata&S1	&\nodata&S1.5	&S1.5 \nl
NGC 3783&\nodata	&S1		&\nodata&S1.2	&\nodata&S1.5	&S1.5 \nl
NGC 4051&\nodata	&S1		&\nodata&S1.5	&\nodata&NLS1	&NLS1 \nl
NGC 4151&\nodata	&S1.5		&\nodata&S1.5	&\nodata&S1.5	&S1.5 \nl
NGC 4235&\nodata	&S1.5		&\nodata&S1.2	&\nodata&S1.2	&S1.5 \nl
NGC 4395&\nodata	&\nodata	&\nodata&\nodata&\nodata&S1.8	&S1.8 \nl
NGC 4507&\nodata	&S2		&\nodata&S2	&\nodata&S1.9	&S1.9 \nl
NGC 4593&\nodata	&\nodata	&\nodata&S1	&\nodata&S1	&BLS1 \nl
NGC 5033&\nodata	&S1.9		&\nodata&\nodata&\nodata&S1.9	&S1.9 \nl
NGC 5252&\nodata	&\nodata	&\nodata&S2	&S2	&S1.9	&S1.9 \nl
NGC 5273&\nodata	&\nodata	&\nodata&S1.9	&\nodata&S1.9	&S1.9\tablenotemark{g} \nl
NGC 5506&\nodata	&S2		&\nodata&S2	&\nodata&S2$_{\rm NIR-BLR}$	&S2$_{\rm NIR-BLR}$ \nl
NGC 5548&\nodata	&S1.5		&\nodata&S1.2	&\nodata&S1.5	&S1.5 \nl
NGC 5674&\nodata	&\nodata	&\nodata&\nodata&S2	&S1.9	&S1.9 \nl
NGC 5929&\nodata	&S2		&\nodata&S2	&\nodata&LINER	&S2 \nl
NGC 7213&\nodata	&\nodata	&\nodata&\nodata&\nodata&S1.5	&S1.5 \nl
NGC 7314&\nodata	&\nodata	&\nodata&S1.9	&\nodata&S1.9	&S1.9 \nl
NGC 7469&\nodata	&S1.5		&\nodata&S1.2	&S1	&S1.5	&S1.5 \nl
NGC 7674&\nodata	&\nodata	&\nodata&\nodata&S2	&S2$_{\rm HBLR}$	&S2$_{\rm HBLR}$ \nl
Mrk 1	&NGC 449	&S2		&\nodata&S2	&S2	&S2	&S2 \nl
Mrk 3	&UGC 3426	&S2		&\nodata&S2	&\nodata&S2$_{\rm HBLR}$	&S2$_{\rm HBLR}$ \nl
Mrk 6	&IC 450		&S1.5		&\nodata&S1.5	&\nodata&S1.5	&S1.5 \nl
Mrk 9	&\nodata	&S1		&\nodata&S1	&\nodata&S1.5	&S1.5 \nl
Mrk 34	&\nodata	&S2		&\nodata&S2	&\nodata&S2	&S2 \nl
Mrk 40	&Arp 151	&S1		&\nodata&S1	&\nodata&S1	&BLS1 \nl
Mrk 42	&\nodata	&S1		&\nodata&S1	&\nodata&NLS1	&NLS1 \nl
Mrk 78	&\nodata	&S2		&\nodata&S2	&\nodata&S2	&S2 \nl
Mrk 79	&UGC 3973	&S1.5		&\nodata&S1.2	&\nodata&S1.2	&S1.5 \nl
Mrk 106	&\nodata	&\nodata	&\nodata&\nodata&\nodata&S1	&BLS1 \nl
Mrk 110	&PG 0921+525	&S1		&\nodata&S1	&S1	&S1.5	&S1.5 \nl
Mrk 142	&PG 1022+519	&S1		&\nodata&S1	&\nodata&S1	&BLS1 \nl
Mrk 176	&UGC 6527	&S2		&\nodata&S2	&\nodata&S2$_{\rm NIR-BLR}$	&S2$_{\rm NIR-BLR}$ \nl
Mrk 268	&\nodata	&S2		&\nodata&S2	&\nodata&S2	&S2 \nl
Mrk 270	&NGC 5283	&S2		&\nodata&S2	&\nodata&S2	&S2 \nl
Mrk 279	&UGC 8823	&S1.5		&\nodata&S1.2	&\nodata&S1	&S1.5 \nl
Mrk 290	&PG 1534+580	&S1.5		&\nodata&S1	&\nodata&S1.5	&S1.5 \nl
Mrk 334	&UGC 6		&S1.8		&\nodata&\nodata&S1.8	&S1.8	&S1.8 \nl
Mrk 335	&PG 0003+199	&S1		&\nodata&S1	&S1	&S1.2	&NLS1\tablenotemark{h} \nl
Mrk 348	&NGC 262	&S2		&\nodata&S2	&S2	&S2$_{\rm HBLR}$	&S2$_{\rm HBLR}$ \nl
Mrk 358	&\nodata	&S1		&\nodata&S1	&\nodata&S1	&BLS1 \nl
Mrk 359	&UGC 1032	&S1.5		&\nodata&S1.5	&S1	&NLS1	&NLS1 \nl
Mrk 374	&\nodata	&S1.5		&\nodata&\nodata&\nodata&S1.2	&S1.5 \nl
Mrk 376	&\nodata	&S1		&\nodata&\nodata&\nodata&S1.5	&S1.5 \nl
Mrk 477	&I Zw 92	&S2		&\nodata&S2	&\nodata&S2$_{\rm HBLR}$	&S2$_{\rm HBLR}$ \nl
Mrk 486	&PG 1535+547	&S1		&\nodata&S1	&\nodata&S1	&BLS1 \nl
Mrk 506	&\nodata	&S1.5		&\nodata&S1.2	&\nodata&S1.5	&S1.5 \nl
Mrk 509	&\nodata	&S1.5		&\nodata&S1.2	&S1	&S1.5	&S1.5 \nl
Mrk 541	&\nodata	&\nodata	&\nodata&\nodata&\nodata&S1	&BLS1 \nl
Mrk 573	&UGC 1214	&S2		&\nodata&S2	&S2	&S2	&S2 \nl
Mrk 607	&NGC 1320	&S2		&\nodata&S2	&\nodata&S2	&S2 \nl
Mrk 618	&\nodata	&S1		&\nodata&S1	&\nodata&S1	&BLS1 \nl
Mrk 686	&NGC 5695	&S2		&\nodata&\nodata&S2	&S2	&S2 \nl
Mrk 699	&III Zw 77	&S1		&\nodata&S1.2	&S1	&S1.5	&S1.5 \nl
Mrk 704	&\nodata	&S1.5		&\nodata&S1.2	&S1	&S1.2	&S1.5 \nl
Mrk 705	&UGC 5025	&S1		&\nodata&\nodata&S1	&S1.2	&BLS1 \nl
Mrk 766	&NGC 4253	&S1.5		&\nodata&S1.5	&S1	&S1.5	&NLS1\tablenotemark{i} \nl
Mrk 783	&\nodata	&S1		&\nodata&\nodata&\nodata&NLS1	&NLS1 \nl
Mrk 817	&UGC 9412	&S1.5		&\nodata&S1.2	&\nodata&S1.5	&S1.5 \nl
Mrk 841	&PG 1501+106	&S1.5		&\nodata&\nodata&\nodata&S1.5	&S1.5 \nl
Mrk 864	&\nodata	&\nodata	&BLS1	&\nodata&\nodata&S1.5	&S1.5 \nl
Mrk 871	&IC 1198	&S1.5		&\nodata&S1.2	&\nodata&S1.5	&S1.5 \nl
Mrk 876	&PG 1613+658	&S1.5		&\nodata&\nodata&\nodata&S1	&S1.5 \nl
Mrk 926	&MCG -2-58-22	&S1.5		&\nodata&S1.2	&\nodata&S1.5	&S1.5 \nl
Mrk 975	&UGC 774	&S1		&\nodata&S1.2	&S1	&S1	&BLS1 \nl
Mrk 993	&UGC 987	&S2		&\nodata&\nodata&\nodata&S1.5	&S1.5 \nl
Mrk 1040&NGC 931	&S1.5		&\nodata&S1.2	&S1	&S1	&S1.5 \nl
Mrk 1126&NGC 7450	&S1.5		&\nodata&S1.5	&\nodata&S1.5	&NLS1\tablenotemark{i} \nl
Mrk 1157&NGC 591	&S2		&\nodata&S2	&\nodata&S2	&S2 \nl
Mrk 1239&\nodata	&S1		&\nodata&S1.2	&S1	&NLS1	&NLS1 \nl
Mrk 1388&\nodata	&S2		&\nodata&S2	&\nodata&S1.9	&S1.9 \nl
Mrk 1393&Tol 1506.3-00	&\nodata	&\nodata&\nodata&\nodata&S1.5	&S1.5 \nl
1H 2107-097	&H 2106-099	&\nodata&\nodata&\nodata&\nodata&S1.2	&BLS1 \nl
2E 1519+2754	&[HB89] 1519+279&\nodata&BLS1	&\nodata&\nodata&S1.2	&S1 \nl
2E 1530+1511	&[HB89] 1530+151&\nodata&BLS1	&\nodata&\nodata&S1.2	&S1 \nl
2E 1556+2725	&PGC 56527	&\nodata&BLS1	&\nodata&\nodata&S1.2	&S1 \nl
I Zw 1		&UGC 545	&S1.5	&\nodata&\nodata&S1	&NLS1	&NLS1 \nl
II Zw 136	&UGC 11763	&S1	&\nodata&S1	&\nodata&S1.5	&S1.5 \nl
Akn 120		&UGC 3271	&\nodata&\nodata&S1	&\nodata&S1	&BLS1 \nl
Akn 564		&UGC 12163	&S1	&\nodata&S1.2	&S1	&NLS1	&NLS1 \nl
Circinus	&\nodata	&\nodata&\nodata&\nodata&\nodata&S2$_{\rm HBLR}$	&S2$_{\rm HBLR}$ \nl
ESO 141-G55	&\nodata	&\nodata&\nodata&S1	&\nodata&S1.2	&BLS1 \nl
ESO 362-G18	&MCG -5-13-17	&\nodata&\nodata&\nodata&\nodata&S1.5	&S1.5 \nl
ESO 439-G09	&Tololo 16	&\nodata&\nodata&\nodata&\nodata&S2	&S2 \nl
Fairall 9	&\nodata	&\nodata&\nodata&S1	&\nodata&S1.2	&BLS1 \nl
Fairall 51	&ESO 140-G043	&\nodata&\nodata&S1.2	&\nodata&S1.5	&S1.5 \nl
Fairall 1116	&\nodata	&\nodata&\nodata&\nodata&\nodata&S1	&BLS1 \nl
IC 4329A	&\nodata	&\nodata&\nodata&S1	&\nodata&S1.2	&BLS1 \nl
KAZ320	&2MASX1 J2259329+245505	&\nodata&\nodata&\nodata&\nodata&NLS1	&NLS1 \nl
MCG -6-30-15	&ESO 383-G035	&\nodata&\nodata&\nodata&\nodata&S1.5	&S1.5 \nl
MCG 8-11-11	&UGC 3374	&S1.5	&\nodata&S1.5	&\nodata&S1.5	&S1.5 \nl
MS 01119-0132	&[HB89] 0111-015&\nodata&NLS1	&\nodata&\nodata&NLS1	&NLS1 \nl
MS 04124-0802	&IRAS 04124-0803&\nodata&S1.5	&\nodata&\nodata&S1.5	&S1.5 \nl
MS 08495+0805	&\nodata	&\nodata&BLS1	&\nodata&\nodata&S1.2	&BLS1 \nl
MS 13285+3135	&[HB89] 1328+315&\nodata&S1.5	&\nodata&\nodata&S1.5	&S1.5 \nl
PKS 2048-57	&IC 5063	&\nodata&\nodata&S2	&\nodata&S1.9	&S1.9 \nl
SBS 1318+605	&\nodata	&\nodata&\nodata&\nodata&\nodata&S1.5	&S1.5 \nl
Tololo 1351-375	&Tololo 113	&\nodata&\nodata&S1.9	&\nodata&S1.9	&S1.9 \nl
Tololo 20	&\nodata	&\nodata&\nodata&\nodata&\nodata&\nodata&BLS1\tablenotemark{j} \nl
Ton 1542	&Akn 374	&\nodata&\nodata&\nodata&\nodata&S1	&BLS1 \nl
UGC 1395	&\nodata	&\nodata&\nodata&S1.8	&\nodata&S1.9	&S1.8 \nl
UGC 6100	&A 1058+45	&\nodata&\nodata&\nodata&S2	&S2	&S2 \nl
UGC 8621	&\nodata	&\nodata&\nodata&\nodata&\nodata&S1.8	&S1.8 \nl
UGC 10683B	&\nodata	&\nodata&\nodata&\nodata&\nodata&S1.5	&S1.5 \nl
UGC 12138	&A 2237+07	&\nodata&\nodata&\nodata&S1	&S1.8	&S1.8\tablenotemark{g} \nl
Zw 0033+45	&CGCG 535-012	&\nodata&\nodata&\nodata&S1	&S1.2	&BLS1 \nl
\cutinhead{Radio Loud AGN}
3C 33           & \nodata    & \nodata & \nodata & \nodata & \nodata & S2       & S2 \nl
3C 120          & II Zw 14   & \nodata & \nodata & \nodata & \nodata & S1.5     & S1.5 \nl
3C 184.1        & \nodata    & \nodata & \nodata & \nodata & \nodata & S2$_{\rm NIR-BLR}$     & S2$_{\rm NIR-BLR}$ \nl
3C 223          & \nodata    & \nodata & \nodata & \nodata & \nodata & S2$_{\rm NIR-BLR}$     & S2$_{\rm NIR-BLR}$ \nl
3C 223.1        & \nodata    & \nodata & \nodata & \nodata & \nodata & S2       & S2 \nl
3C 327       &IRAS 15599+0206& \nodata & \nodata & \nodata & \nodata & S2       & S2 \nl
3C 390.3        & \nodata    & \nodata & \nodata & \nodata & \nodata & S1.5     & S1.5 \nl
3C 445      &IRAS F22212-0221& \nodata & \nodata & \nodata & \nodata & S1.5     & S1.5 \nl
3C 452          & \nodata    & \nodata & \nodata & \nodata & \nodata & S2       & S2 \\
\enddata
\tablenotetext{a}{Dahari \& De Robertis (1988)}
\tablenotetext{b}{Stephens (1989)}
\tablenotetext{c}{Whittle (1992)}
\tablenotetext{d}{Cruz-Gonz$\acute{\rm a}$lez et al. (1994)}
\tablenotetext{e}{V$\acute{\rm e}$ron-Cetty \& V$\acute{\rm e}$ron (1998)}
\tablenotetext{f}{DDR88 mentioned that these galaxies show marginal properties
                  between the S2 and the starburst galaxy.}
\tablenotetext{g}{Though there is discrepancy in the classification of these galaxies
                  among the references, we classify these object
                  following more widespread classification
                  (see e.g. the NED; NASA extragalactic database).}
\tablenotetext{h}{This galaxy is classified following Vaughan et al. (1999).}
\tablenotetext{i}{These galaxies are classified following Osterbrock \& Pogge (1985)
                  and Boller et al. (1996).}
\tablenotetext{j}{This galaxy is classified following the classification of the NED.}
\end{deluxetable}
%%%%%%%%%% TABLE 2 %%%%%%%%%%

%%%%%%%%%% TABLE 3 %%%%%%%%%%
\begin{deluxetable}{ll}
\tablenum{3}
\tablecaption{References for Optical Emission-Lines \label{tbl-3}}
\tablewidth{0pt}
\tablehead{
\colhead{Abbreviation} &
\colhead{References}
}
\startdata
C77 & Costero \& Osterbrock (1977) \nl
C81 & Cohen \& Osterbrock (1981) \nl
C83 & Cohen (1983)  \nl
C91 & Crenshaw, Peterson, Korista, Wagner, \& Aufdenberg (1991) \nl
C94 & Cruz-Gonzalez, Carrasco, Serrano, Guichard, Dultzin-Hacyan, \& Bisiacchi (1994) \nl
D78 & Davidson \& Kinman (1978) \nl
D86 & De Robertis \& Osterbrock (1986) \nl
D88 & Diaz, Prieto, \& Wamsteker (1988) \nl
E97 & Erkens, Wagner, \& Appenzeller (1997) \nl
E99 & Erkens, Wagner, \& Appenzeller, private communication (1999) \nl
F83 & Fosbury \& Sansom (1983) \nl
K78 & Koski (1978) \nl
K79 & Kunth \& Sargent (1979) \nl
K94 & Kraemer, Wu, Crenshaw, \& Harrington (1994) \nl
M88 & Morris \& Ward (1988) \nl
M94 & Martel \& Osterbrock (1994) \nl
M95 & Murayama (1995) \nl
M98 & Murayama, Taniguchi, \& Iwasawa (1998) \nl 
O77 & Osterbrock (1977) \nl 
O78 & O'Connell \& Kingham (1978) \nl
O81 & Osterbrock (1981a) \nl
O85 & Osterbrock (1985)  \nl
O93 & Osterbrock \& Martel (1993) \nl
O94 & Oliva, Salvati, Moorwood, \& Marconi (1994) \nl
OP  & Osterbrock \& Pogge (1985) \nl
P78 & Phillips (1978) \nl
P84 & Penston, Fosbury, Boksenberg, Ward, \& Wilson (1984) \nl
R97 & Reynolds, Ward, Fabian, \& Celotti (1997) \nl
S80 & Shuder (1980) \nl
S89 & Stephens (1989) \nl
V88 & Veilleux (1988) \nl
W92 & Winkler (1992) \nl
Z92 & Zamorano, Gallego, Rego, Vitores, \& Gonzalez-Riestra (1992) \\
\enddata
\end{deluxetable}
%%%%%%%%%% TABLE 3 %%%%%%%%%%

%%%%%%%%%% TABLE 4 %%%%%%%%%%
\begin{deluxetable}{lcccc}
\tablewidth{0pt}
\tablenum{4}
\tablecaption{The Detection Rates of the High-Ionization
              Emission Lines \label{tbl-4}}
\tablehead{
\colhead{Class} &
\colhead{[Fe VII]$\lambda$6087} &
\colhead{[Fe X]$\lambda$6374} &
\colhead{[Fe XI]$\lambda$7892} &
\colhead{In Total\tablenotemark{a}}
}
\startdata
NLS1     & 11/30 & 10/29 & 5/8  & 12/31 (38.7\%) \nl
BLS1     & 21/58 & 13/52 & 5/19 & 24/58 (41.4\%) \nl
S1.5     & 38/67 & 31/62 & 8/30 & 43/67 (64.2\%) \nl
S2$^{+}$ & 19/31 & 23/29 & 2/9  & 26/31 (83.9\%) \nl
S2$^{-}$ & 15/39 &  9/35 & 2/7  & 19/40 (47.5\%) \\
\enddata
\tablenotetext{a}{The objects in which any high-ionization emission line
                  is detected are counted in these numbers.}
\end{deluxetable}
%%%%%%%%%% TABLE 4 %%%%%%%%%%

%%%%%%%%%% TABLE 7 %%%%%%%%%%
\begin{deluxetable}{lcccccc}
\tablenum{7}
\tablecaption{The Results of the KS Test 
              Concerning the Redshift \label{tbl-7}}
\tablewidth{0pt}
\tablehead{
\colhead{Class\tablenotemark{a}} &
\colhead{NLS1} &
\colhead{BLS1} & 
\colhead{S1.5} & 
\colhead{S2$_{\rm total}$\tablenotemark{b}} &
\colhead{S2$^+$} &
\colhead{S2$^-$}
}
\startdata
NLS1             &\nodata& 1.264$\times$10$^{-2}$ & 5.466$\times$10$^{-1}$ & 4.200$\times$10$^{-1}$ & 3.539$\times$10$^{-1}$ & 8.035$\times$10$^{-1}$ \nl
                 &\nodata& 1.264$\times$10$^{-2}$ & 7.243$\times$10$^{-1}$ & 1.641$\times$10$^{-1}$ & 2.182$\times$10$^{-1}$ & 3.146$\times$10$^{-1}$ \nl
BLS1             &\nodata& \nodata                & 4.103$\times$10$^{-2}$ & 5.227$\times$10$^{-5}$ & 8.097$\times$10$^{-6}$ & 3.972$\times$10$^{-2}$ \nl
                 &\nodata& \nodata                & 2.631$\times$10$^{-2}$ & 1.104$\times$10$^{-6}$ & 9.773$\times$10$^{-7}$ & 2.183$\times$10$^{-3}$ \nl
S1.5             &\nodata& \nodata                & \nodata                & 1.568$\times$10$^{-2}$ & 2.502$\times$10$^{-3}$ & 6.824$\times$10$^{-1}$ \nl
                 &\nodata& \nodata                & \nodata                & 2.182$\times$10$^{-3}$ & 1.325$\times$10$^{-3}$ & 1.603$\times$10$^{-1}$ \nl
S2$_{\rm total}$\tablenotemark{b} &\nodata& \nodata                & \nodata                & \nodata                & \nodata                & \nodata                \nl
                 &\nodata& \nodata                & \nodata                & \nodata                & \nodata                & \nodata                \nl
S2$^+$           &\nodata& \nodata                & \nodata                & \nodata                & \nodata                & 1.330$\times$10$^{-1}$ \nl
                 &\nodata& \nodata                & \nodata                & \nodata                & \nodata                & 1.366$\times$10$^{-1}$ \nl
S2$^-$           &\nodata& \nodata                & \nodata                & \nodata                & \nodata                & \nodata                \nl
                 &\nodata& \nodata                & \nodata                & \nodata                & \nodata                & \nodata                \nl
\enddata
\tablenotetext{a}{The upper line for each class gives the KS probabilities
                  in the case of including the radio-loud objects, and
                  the lower lines give those in the case of excluding the
                  radio-loud objects.}
\tablenotetext{b}{``S2$_{\rm total}$'' means ``S2$^+$'' plus ``S2$^-$''.}
\end{deluxetable}
%%%%%%%%%% TABLE 7 %%%%%%%%%%

%%%%%%%%%% TABLE 8 %%%%%%%%%%
\begin{deluxetable}{lcccccc}
\tablenum{8}
\tablecaption{The Results of the KS Test 
              Concerning the 60$\mu$m Luminosities of Our Sample \label{tbl-8}}
\tablewidth{0pt}
\tablehead{
\colhead{Class\tablenotemark{a}} &
\colhead{NLS1s} &
\colhead{BLS1s} & 
\colhead{S1.5s} & 
\colhead{S2$_{\rm total}$s\tablenotemark{b}} &
\colhead{S2$^+$s} &
\colhead{S2$^-$s}
}
\startdata
NLS1s             &\nodata& 4.494$\times$10$^{-1}$ & 7.662$\times$10$^{-1}$ & 9.992$\times$10$^{-1}$ & 9.937$\times$10$^{-1}$ & 9.427$\times$10$^{-1}$ \nl
                  &\nodata& 4.494$\times$10$^{-1}$ & 8.571$\times$10$^{-1}$ & 9.998$\times$10$^{-1}$ & 9.937$\times$10$^{-1}$ & 9.783$\times$10$^{-1}$ \nl
BLS1s             &\nodata& \nodata                & 2.175$\times$10$^{-1}$ & 1.260$\times$10$^{-1}$ & 2.294$\times$10$^{-1}$ & 1.661$\times$10$^{-1}$ \nl
                  &\nodata& \nodata                & 1.306$\times$10$^{-1}$ & 1.037$\times$10$^{-1}$ & 2.294$\times$10$^{-1}$ & 1.072$\times$10$^{-1}$ \nl
S1.5s             &\nodata& \nodata                & \nodata                & 7.097$\times$10$^{-1}$ & 7.969$\times$10$^{-1}$ & 7.049$\times$10$^{-1}$ \nl
                  &\nodata& \nodata                & \nodata                & 7.304$\times$10$^{-1}$ & 9.088$\times$10$^{-1}$ & 6.348$\times$10$^{-1}$ \nl
S2$_{\rm total}$s &\nodata& \nodata                & \nodata                & \nodata                & \nodata                & \nodata                \nl
                  &\nodata& \nodata                & \nodata                & \nodata                & \nodata                & \nodata                \nl
S2$^+$s           &\nodata& \nodata                & \nodata                & \nodata                & \nodata                & 5.086$\times$10$^{-1}$ \nl
                  &\nodata& \nodata                & \nodata                & \nodata                & \nodata                & 6.010$\times$10$^{-1}$ \nl
S2$^-$s           &\nodata& \nodata                & \nodata                & \nodata                & \nodata                & \nodata                \nl
                  &\nodata& \nodata                & \nodata                & \nodata                & \nodata                & \nodata                \nl
\enddata
\tablenotetext{a}{The upper line for each class gives the KS probabilities
                  in the case of including the radio-loud objects, and
                  the lower lines give those in the case of excluding the
                  radio-loud objects.}
\end{deluxetable}
%%%%%%%%%% TABLE 8 %%%%%%%%%%

%%%%%%%%%% TABLE 9 %%%%%%%%%%
\begin{deluxetable}{lcccccc}
\tablenum{9}
\tablecaption{The Results of the KS Test 
              Concirning the Luminosities of the Low-Ionization 
              Emission Lines\label{tbl-9}}
\tablewidth{0pt}
\tablehead{
\colhead{Class\tablenotemark{a}} &
\colhead{NLS1} &
\colhead{BLS1} & 
\colhead{S1.5} & 
\colhead{S2$_{\rm total}$\tablenotemark{b}} &
\colhead{S2$^+$} &
\colhead{S2$^-$}
}
\startdata
\cutinhead{{\it L}$_{\rm [O III]}\lambda$5007}
NLS1             &\nodata& 5.263$\times$10$^{-1}$ & 7.513$\times$10$^{-1}$ & 8.837$\times$10$^{-1}$ & 9.158$\times$10$^{-1}$ & 6.220$\times$10$^{-1}$ \nl
                 &\nodata& 5.263$\times$10$^{-1}$ & 7.779$\times$10$^{-1}$ & 9.919$\times$10$^{-1}$ &9.9999$\times$10$^{-1}$ & 7.864$\times$10$^{-1}$ \nl
BLS1             &\nodata& \nodata                & 5.188$\times$10$^{-1}$ & 6.077$\times$10$^{-1}$ & 6.827$\times$10$^{-1}$ & 6.325$\times$10$^{-1}$ \nl
                 &\nodata& \nodata                & 5.805$\times$10$^{-1}$ & 3.852$\times$10$^{-1}$ & 4.535$\times$10$^{-1}$ & 4.370$\times$10$^{-1}$ \nl
S1.5             &\nodata& \nodata                & \nodata                & 9.636$\times$10$^{-1}$ & 8.272$\times$10$^{-1}$ & 9.020$\times$10$^{-1}$ \nl
                 &\nodata& \nodata                & \nodata                & 7.769$\times$10$^{-1}$ & 7.200$\times$10$^{-1}$ & 8.878$\times$10$^{-1}$ \nl
S2$_{\rm total}$\tablenotemark{b} &\nodata& \nodata                & \nodata                & \nodata                & \nodata                & \nodata                \nl
                 &\nodata& \nodata                & \nodata                & \nodata                & \nodata                & \nodata                \nl
S2$^+$           &\nodata& \nodata                & \nodata                & \nodata                & \nodata                & 9.160$\times$10$^{-1}$ \nl
                 &\nodata& \nodata                & \nodata                & \nodata                & \nodata                & 9.524$\times$10$^{-1}$ \nl
S2$^-$           &\nodata& \nodata                & \nodata                & \nodata                & \nodata                & \nodata                \nl
                 &\nodata& \nodata                & \nodata                & \nodata                & \nodata                & \nodata                \nl
\cutinhead{{\it L}$_{\rm [S II]}\lambda\lambda$6717,6731}
NLS1             &\nodata& 1.822$\times$10$^{-1}$ & 2.596$\times$10$^{-1}$ & 5.309$\times$10$^{-2}$ & 1.283$\times$10$^{-1}$ & 6.558$\times$10$^{-2}$ \nl
                 &\nodata& 1.822$\times$10$^{-1}$ & 3.425$\times$10$^{-1}$ & 2.086$\times$10$^{-1}$ & 2.108$\times$10$^{-1}$ & 4.339$\times$10$^{-1}$ \nl
BLS1             &\nodata& \nodata                & 6.666$\times$10$^{-1}$ & 9.452$\times$10$^{-1}$ & 9.892$\times$10$^{-1}$ & 9.495$\times$10$^{-1}$ \nl
                 &\nodata& \nodata                & 6.214$\times$10$^{-1}$ & 9.747$\times$10$^{-1}$ & 9.855$\times$10$^{-1}$ & 9.852$\times$10$^{-1}$ \nl
S1.5             &\nodata& \nodata                & \nodata                & 7.806$\times$10$^{-1}$ & 8.857$\times$10$^{-1}$ & 6.350$\times$10$^{-1}$ \nl
                 &\nodata& \nodata                & \nodata                & 8.581$\times$10$^{-1}$ & 8.800$\times$10$^{-1}$ & 9.830$\times$10$^{-1}$ \nl
S2$_{\rm total}$\tablenotemark{b} &\nodata& \nodata                & \nodata                & \nodata                & \nodata                & \nodata                \nl
                 &\nodata& \nodata                & \nodata                & \nodata                & \nodata                & \nodata                \nl
S2$^+$           &\nodata& \nodata                & \nodata                & \nodata                & \nodata                & 9.889$\times$10$^{-1}$ \nl
                 &\nodata& \nodata                & \nodata                & \nodata                & \nodata                & 9.897$\times$10$^{-1}$ \nl
S2$^-$           &\nodata& \nodata                & \nodata                & \nodata                & \nodata                & \nodata                \nl
                 &\nodata& \nodata                & \nodata                & \nodata                & \nodata                & \nodata                \nl
\cutinhead{{\it L}$_{\rm [O I]}\lambda$6300}
NLS1             &\nodata& 4.384$\times$10$^{-1}$ & 3.608$\times$10$^{-2}$ & 1.316$\times$10$^{-1}$ & 2.555$\times$10$^{-1}$ & 1.535$\times$10$^{-1}$ \nl
                 &\nodata& 4.384$\times$10$^{-1}$ & 4.299$\times$10$^{-2}$ & 3.079$\times$10$^{-1}$ & 3.889$\times$10$^{-1}$ & 5.019$\times$10$^{-1}$ \nl
BLS1             &\nodata& \nodata                & 8.700$\times$10$^{-1}$ & 9.197$\times$10$^{-1}$ & 8.719$\times$10$^{-1}$ & 8.361$\times$10$^{-1}$ \nl
                 &\nodata& \nodata                & 8.034$\times$10$^{-1}$ & 6.882$\times$10$^{-1}$ & 7.471$\times$10$^{-1}$ & 8.581$\times$10$^{-1}$ \nl
S1.5             &\nodata& \nodata                & \nodata                & 7.326$\times$10$^{-1}$ & 9.343$\times$10$^{-1}$ & 8.019$\times$10$^{-1}$ \nl
                 &\nodata& \nodata                & \nodata                & 6.632$\times$10$^{-1}$ & 9.268$\times$10$^{-1}$ & 6.851$\times$10$^{-1}$ \nl
S2$_{\rm total}$\tablenotemark{b} &\nodata& \nodata                & \nodata                & \nodata                & \nodata                & \nodata                \nl
                 &\nodata& \nodata                & \nodata                & \nodata                & \nodata                & \nodata                \nl
S2$^+$           &\nodata& \nodata                & \nodata                & \nodata                & \nodata                & 9.379$\times$10$^{-1}$ \nl
                 &\nodata& \nodata                & \nodata                & \nodata                & \nodata                & 9.665$\times$10$^{-1}$ \nl
S2$^-$           &\nodata& \nodata                & \nodata                & \nodata                & \nodata                & \nodata                \nl
                 &\nodata& \nodata                & \nodata                & \nodata                & \nodata                & \nodata                \nl
\enddata
\tablenotetext{a}{The upper line for each class gives the KS probabilities
                  in the case of including the radio-loud objects, and
                  the lower lines give those in the case of excluding the
                  radio-loud objects.}
\tablenotetext{b}{``S2$_{\rm total}$'' means ``S2$^+$'' plus ``S2$^-$''.}
\end{deluxetable}
%%%%%%%%%% TABLE 9 %%%%%%%%%%

%%%%%%%%%% TABLE 10 %%%%%%%%%%
\begin{deluxetable}{lcccccc}
\tablenum{10}
\tablecaption{The Results of the KS Test 
              Concerning [Fe {\sc vii}] \label{tbl-10}}
\tablewidth{0pt}
\tablehead{
\colhead{Class\tablenotemark{a}} &
\colhead{NLS1} &
\colhead{BLS1} & 
\colhead{S1.5} & 
\colhead{S2$_{\rm total}$\tablenotemark{b}} &
\colhead{S2$^+$} &
\colhead{S2$^-$}
}
\startdata
\cutinhead{[Fe {\sc vii}]$\lambda$6087 / [O {\sc iii}]$\lambda$5007}
NLS1             &\nodata& 7.954$\times$10$^{-2}$ & 3.460$\times$10$^{-1}$ & 1.050$\times$10$^{-4}$ & 5.309$\times$10$^{-4}$ & 8.114$\times$10$^{-4}$ \nl
                 &\nodata& 7.954$\times$10$^{-2}$ & 5.626$\times$10$^{-1}$ & 2.253$\times$10$^{-4}$ & 6.874$\times$10$^{-4}$ & 3.954$\times$10$^{-3}$ \nl
BLS1             &\nodata& \nodata                & 2.150$\times$10$^{-3}$ & 1.387$\times$10$^{-8}$ & 9.316$\times$10$^{-7}$ & 2.190$\times$10$^{-6}$ \nl
                 &\nodata& \nodata                & 5.622$\times$10$^{-3}$ & 8.041$\times$10$^{-8}$ & 1.609$\times$10$^{-6}$ & 2.644$\times$10$^{-5}$ \nl
S1.5             &\nodata& \nodata                & \nodata                & 5.093$\times$10$^{-6}$ & 1.350$\times$10$^{-5}$ & 9.136$\times$10$^{-4}$ \nl
                 &\nodata& \nodata                & \nodata                & 1.805$\times$10$^{-5}$ & 2.375$\times$10$^{-4}$ & 2.417$\times$10$^{-3}$ \nl
S2$_{\rm total}$\tablenotemark{b} &\nodata& \nodata                & \nodata                & \nodata                & \nodata                & \nodata                \nl
                 &\nodata& \nodata                & \nodata                & \nodata                & \nodata                & \nodata                \nl
S2$^+$           &\nodata& \nodata                & \nodata                & \nodata                & \nodata                & 5.090$\times$10$^{-1}$ \nl
                 &\nodata& \nodata                & \nodata                & \nodata                & \nodata                & 9.630$\times$10$^{-1}$ \nl
S2$^-$           &\nodata& \nodata                & \nodata                & \nodata                & \nodata                & \nodata                \nl
                 &\nodata& \nodata                & \nodata                & \nodata                & \nodata                & \nodata                \nl
\cutinhead{[Fe {\sc vii}]$\lambda$6087 / [S {\sc ii}]$\lambda\lambda$6717,6731}
NLS1             &\nodata& 1.355$\times$10$^{-1}$ & 9.684$\times$10$^{-1}$ & 5.305$\times$10$^{-4}$ & 3.406$\times$10$^{-3}$ & 1.699$\times$10$^{-3}$ \nl
                 &\nodata& 1.355$\times$10$^{-1}$ & 9.756$\times$10$^{-1}$ & 1.191$\times$10$^{-3}$ & 4.596$\times$10$^{-3}$ & 6.899$\times$10$^{-3}$ \nl
BLS1             &\nodata& \nodata                & 1.388$\times$10$^{-1}$ & 4.616$\times$10$^{-7}$ & 2.248$\times$10$^{-5}$ & 1.130$\times$10$^{-5}$ \nl
                 &\nodata& \nodata                & 1.930$\times$10$^{-1}$ & 2.468$\times$10$^{-6}$ & 4.265$\times$10$^{-5}$ & 1.507$\times$10$^{-4}$ \nl
S1.5             &\nodata& \nodata                & \nodata                & 7.373$\times$10$^{-5}$ & 3.320$\times$10$^{-3}$ & 1.586$\times$10$^{-3}$ \nl
                 &\nodata& \nodata                & \nodata                & 5.509$\times$10$^{-4}$ & 6.479$\times$10$^{-3}$ & 9.612$\times$10$^{-3}$ \nl
S2$_{\rm total}$\tablenotemark{b} &\nodata& \nodata                & \nodata                & \nodata                & \nodata                & \nodata                \nl
                 &\nodata& \nodata                & \nodata                & \nodata                & \nodata                & \nodata                \nl
S2$^+$           &\nodata& \nodata                & \nodata                & \nodata                & \nodata                & 9.302$\times$10$^{-1}$ \nl
                 &\nodata& \nodata                & \nodata                & \nodata                & \nodata                & 9.474$\times$10$^{-1}$ \nl
S2$^-$           &\nodata& \nodata                & \nodata                & \nodata                & \nodata                & \nodata                \nl
                 &\nodata& \nodata                & \nodata                & \nodata                & \nodata                & \nodata                \nl
\cutinhead{[Fe {\sc vii}]$\lambda$6087 / [O {\sc i}]$\lambda$6300}
NLS1             &\nodata& 1.902$\times$10$^{-1}$ & 9.684$\times$10$^{-1}$ & 2.995$\times$10$^{-2}$ & 1.958$\times$10$^{-2}$ & 1.036$\times$10$^{-1}$ \nl
                 &\nodata& 1.902$\times$10$^{-1}$ & 9.972$\times$10$^{-1}$ & 3.608$\times$10$^{-2}$ & 2.481$\times$10$^{-2}$ & 1.108$\times$10$^{-1}$ \nl
BLS1             &\nodata& \nodata                & 6.286$\times$10$^{-2}$ & 6.120$\times$10$^{-6}$ & 2.519$\times$10$^{-5}$ & 3.125$\times$10$^{-4}$ \nl
                 &\nodata& \nodata                & 8.606$\times$10$^{-2}$ & 6.405$\times$10$^{-6}$ & 3.748$\times$10$^{-5}$ & 4.708$\times$10$^{-4}$ \nl
S1.5             &\nodata& \nodata                & \nodata                & 9.470$\times$10$^{-4}$ & 1.631$\times$10$^{-3}$ & 3.159$\times$10$^{-2}$ \nl
                 &\nodata& \nodata                & \nodata                & 2.561$\times$10$^{-3}$ & 5.638$\times$10$^{-3}$ & 6.327$\times$10$^{-2}$ \nl
S2$_{\rm total}$\tablenotemark{b} &\nodata& \nodata                & \nodata                & \nodata                & \nodata                & \nodata                \nl
                 &\nodata& \nodata                & \nodata                & \nodata                & \nodata                & \nodata                \nl
S2$^+$           &\nodata& \nodata                & \nodata                & \nodata                & \nodata                & 7.579$\times$10$^{-1}$ \nl
                 &\nodata& \nodata                & \nodata                & \nodata                & \nodata                & 7.159$\times$10$^{-1}$ \nl
S2$^-$           &\nodata& \nodata                & \nodata                & \nodata                & \nodata                & \nodata                \nl
                 &\nodata& \nodata                & \nodata                & \nodata                & \nodata                & \nodata                \nl
\enddata
\tablenotetext{a}{The upper line for each class gives the KS probabilities
                  in the case of including the radio-loud objects, and
                  the lower lines give those in the case of excluding the
                  radio-loud objects.}
\tablenotetext{b}{``S2$_{\rm total}$'' means ``S2$^+$'' plus ``S2$^-$''.}
\end{deluxetable}
%%%%%%%%%% TABLE 10 %%%%%%%%%%

%%%%%%%%%% TABLE 11 %%%%%%%%%%
\begin{deluxetable}{lcccccc}
\tablenum{11}
\tablecaption{The Results of the KS Test 
              Concerning [Fe {\sc x}] \label{tbl-11}}
\tablewidth{0pt}
\tablehead{
\colhead{Class\tablenotemark{a}} &
\colhead{NLS1} &
\colhead{BLS1} & 
\colhead{S1.5} & 
\colhead{S2$_{\rm total}$\tablenotemark{b}} &
\colhead{S2$^+$} &
\colhead{S2$^-$}
}
\startdata
\cutinhead{[Fe {\sc x}]$\lambda$6374 / [O {\sc iii}]$\lambda$5007}
NLS1             &\nodata& 3.409$\times$10$^{-1}$ & 9.488$\times$10$^{-2}$ & 1.100$\times$10$^{-2}$ & 1.601$\times$10$^{-2}$ & 4.732$\times$10$^{-2}$ \nl
                 &\nodata& 3.409$\times$10$^{-1}$ & 1.061$\times$10$^{-1}$ & 1.662$\times$10$^{-2}$ & 2.098$\times$10$^{-2}$ & 6.066$\times$10$^{-2}$ \nl
BLS1             &\nodata& \nodata                & 2.938$\times$10$^{-1}$ & 8.839$\times$10$^{-4}$ & 2.091$\times$10$^{-3}$ & 8.193$\times$10$^{-3}$ \nl
                 &\nodata& \nodata                & 3.981$\times$10$^{-1}$ & 1.177$\times$10$^{-3}$ & 3.548$\times$10$^{-3}$ & 1.230$\times$10$^{-2}$ \nl
S1.5             &\nodata& \nodata                & \nodata                & 2.211$\times$10$^{-2}$ & 4.121$\times$10$^{-2}$ & 6.827$\times$10$^{-2}$ \nl
                 &\nodata& \nodata                & \nodata                & 2.509$\times$10$^{-2}$ & 3.932$\times$10$^{-2}$ & 8.088$\times$10$^{-2}$ \nl
S2$_{\rm total}$\tablenotemark{b} &\nodata& \nodata                & \nodata                & \nodata                & \nodata                & \nodata                \nl
                 &\nodata& \nodata                & \nodata                & \nodata                & \nodata                & \nodata                \nl
S2$^+$           &\nodata& \nodata                & \nodata                & \nodata                & \nodata                & 9.700$\times$10$^{-1}$ \nl
                 &\nodata& \nodata                & \nodata                & \nodata                & \nodata                & 9.705$\times$10$^{-1}$ \nl
S2$^-$           &\nodata& \nodata                & \nodata                & \nodata                & \nodata                & \nodata                \nl
                 &\nodata& \nodata                & \nodata                & \nodata                & \nodata                & \nodata                \nl
\cutinhead{[Fe {\sc x}]$\lambda$6374 / [S {\sc ii}]$\lambda\lambda$6717,6731}
NLS1             &\nodata& 3.754$\times$10$^{-1}$ & 4.846$\times$10$^{-1}$ & 1.852$\times$10$^{-3}$ & 3.441$\times$10$^{-3}$ & 1.224$\times$10$^{-2}$ \nl
                 &\nodata& 3.754$\times$10$^{-1}$ & \nodata                & 3.142$\times$10$^{-3}$ & 4.895$\times$10$^{-3}$ & 1.717$\times$10$^{-2}$ \nl
BLS1             &\nodata& \nodata                & 3.323$\times$10$^{-2}$ & 9.567$\times$10$^{-6}$ & 7.510$\times$10$^{-5}$ & 9.464$\times$10$^{-5}$ \nl
                 &\nodata& \nodata                & \nodata                & 1.470$\times$10$^{-5}$ & 1.052$\times$10$^{-4}$ & 1.776$\times$10$^{-4}$ \nl
S1.5             &\nodata& \nodata                & \nodata                & 4.437$\times$10$^{-4}$ & 1.302$\times$10$^{-3}$ & 7.685$\times$10$^{-3}$ \nl
                 &\nodata& \nodata                & \nodata                & 1.223$\times$10$^{-4}$ & 6.629$\times$10$^{-4}$ & 1.279$\times$10$^{-2}$ \nl
S2$_{\rm total}$\tablenotemark{b} &\nodata& \nodata                & \nodata                & \nodata                & \nodata                & \nodata                \nl
                 &\nodata& \nodata                & \nodata                & \nodata                & \nodata                & \nodata                \nl
S2$^+$           &\nodata& \nodata                & \nodata                & \nodata                & \nodata                & 9.563$\times$10$^{-1}$ \nl
                 &\nodata& \nodata                & \nodata                & \nodata                & \nodata                & 9.560$\times$10$^{-1}$ \nl
S2$^-$           &\nodata& \nodata                & \nodata                & \nodata                & \nodata                & \nodata                \nl
                 &\nodata& \nodata                & \nodata                & \nodata                & \nodata                & \nodata                \nl
\cutinhead{[Fe {\sc x}]$\lambda$6374 / [O {\sc i}]$\lambda$6300}
NLS1             &\nodata& 5.472$\times$10$^{-1}$ & 6.452$\times$10$^{-1}$ & 2.938$\times$10$^{-2}$ & 3.121$\times$10$^{-4}$ & 2.470$\times$10$^{-2}$ \nl
                 &\nodata& 5.472$\times$10$^{-1}$ & 7.639$\times$10$^{-1}$ & 4.630$\times$10$^{-4}$ & 4.175$\times$10$^{-4}$ & 5.126$\times$10$^{-2}$ \nl
BLS1             &\nodata& \nodata                & 7.480$\times$10$^{-2}$ & 1.165$\times$10$^{-4}$ & 1.196$\times$10$^{-4}$ & 9.766$\times$10$^{-3}$ \nl
                 &\nodata& \nodata                & 9.307$\times$10$^{-2}$ & 1.802$\times$10$^{-4}$ & 1.583$\times$10$^{-4}$ & 1.342$\times$10$^{-2}$ \nl
S1.5             &\nodata& \nodata                & \nodata                & 1.903$\times$10$^{-3}$ & 2.897$\times$10$^{-3}$ & 9.498$\times$10$^{-2}$ \nl
                 &\nodata& \nodata                & \nodata                & 7.688$\times$10$^{-4}$ & 2.466$\times$10$^{-3}$ & 5.643$\times$10$^{-2}$ \nl
S2$_{\rm total}$\tablenotemark{b} &\nodata& \nodata                & \nodata                & \nodata                & \nodata                & \nodata                \nl
                 &\nodata& \nodata                & \nodata                & \nodata                & \nodata                & \nodata                \nl
S2$^+$           &\nodata& \nodata                & \nodata                & \nodata                & \nodata                & 8.473$\times$10$^{-1}$ \nl
                 &\nodata& \nodata                & \nodata                & \nodata                & \nodata                & 8.489$\times$10$^{-1}$ \nl
S2$^-$           &\nodata& \nodata                & \nodata                & \nodata                & \nodata                & \nodata                \nl
                 &\nodata& \nodata                & \nodata                & \nodata                & \nodata                & \nodata                \\
\enddata
\tablenotetext{a}{The upper line for each class gives the KS probabilities
                  in the case of including the radio-loud objects, and
                  the lower lines give those in the case of excluding the
                  radio-loud objects.}
\tablenotetext{b}{``S2$_{\rm total}$'' means ``S2$^+$'' plus ``S2$^-$''.}
\end{deluxetable}
%%%%%%%%%% TABLE 11 %%%%%%%%%%

%%%%%%%%%% TABLE 12 %%%%%%%%%%
\begin{deluxetable}{lcccccc}
\tablenum{12}
\tablecaption{The Results of the KS Test 
              Concerning [Fe {\sc x}]/[Fe {\sc vii}] \label{tbl-12}}
\tablewidth{0pt}
\tablehead{
\colhead{Class\tablenotemark{a}} &
\colhead{NLS1} &
\colhead{BLS1} & 
\colhead{S1.5} & 
\colhead{S2$_{\rm total}$\tablenotemark{b}} &
\colhead{S2$^+$} &
\colhead{S2$^-$}
}
\startdata
NLS1             &\nodata& 7.488$\times$10$^{-1}$ & 4.456$\times$10$^{-1}$ & 1.821$\times$10$^{-1}$ & 2.976$\times$10$^{-1}$ & 6.258$\times$10$^{-2}$ \nl
                 &\nodata& 7.488$\times$10$^{-1}$ & 5.480$\times$10$^{-1}$ & 1.343$\times$10$^{-1}$ & 2.976$\times$10$^{-1}$ & 9.538$\times$10$^{-2}$ \nl
BLS1             &\nodata& \nodata                & 2.485$\times$10$^{-1}$ & 1.021$\times$10$^{-1}$ & 1.917$\times$10$^{-1}$ & 2.342$\times$10$^{-1}$ \nl
                 &\nodata& \nodata                & 2.774$\times$10$^{-1}$ & 7.193$\times$10$^{-2}$ & 1.917$\times$10$^{-1}$ & 8.776$\times$10$^{-2}$ \nl
S1.5             &\nodata& \nodata                & \nodata                & 4.697$\times$10$^{-1}$ & 5.860$\times$10$^{-1}$ & 2.082$\times$10$^{-1}$ \nl
                 &\nodata& \nodata                & \nodata                & 2.898$\times$10$^{-1}$ & 6.226$\times$10$^{-1}$ & 2.801$\times$10$^{-1}$ \nl
S2$_{\rm total}$\tablenotemark{b} &\nodata& \nodata                & \nodata                & \nodata                & \nodata                & \nodata                \nl
                 &\nodata& \nodata                & \nodata                & \nodata                & \nodata                & \nodata                \nl
S2$^+$           &\nodata& \nodata                & \nodata                & \nodata                & \nodata                & 3.947$\times$10$^{-1}$ \nl
                 &\nodata& \nodata                & \nodata                & \nodata                & \nodata                & 4.898$\times$10$^{-1}$ \nl
S2$^-$           &\nodata& \nodata                & \nodata                & \nodata                & \nodata                & \nodata                \nl
                 &\nodata& \nodata                & \nodata                & \nodata                & \nodata                & \nodata                \nl
\enddata
\tablenotetext{a}{The upper line for each class gives the KS probabilities
                  in the case of including the radio-loud objects, and
                  the lower lines give those in the case of excluding the
                  radio-loud objects.}
\tablenotetext{b}{``S2$_{\rm total}$'' means ``S2$^+$'' plus ``S2$^-$''.}
\end{deluxetable}
%%%%%%%%%% TABLE 12 %%%%%%%%%%

%%%%%%%%%% TABLE 13 %%%%%%%%%%
\begin{deluxetable}{lccc}
\tablenum{13}
\tablecaption{Effects of the Correction for the Dust Extinction to 
              the Line Ratio of HINER Components to 
              Low-Ionization Emission Lines \label{tbl-13}}
\tablewidth{30pc}
\tablehead{
\colhead{} &
\colhead{{\it A}$_{\rm V}$ = 1.0} &
\colhead{{\it A}$_{\rm V}$ = 5.2\tablenotemark{a}} &
\colhead{{\it A}$_{\rm V}$ = 10.0}
}
\startdata
[Fe {\sc vii}]6087/[O {\sc iii}]5007     & --0.091 & --0.474 & --0.911 \nl
[Fe {\sc vii}]6087/[S {\sc ii}]6717,6731 &   0.040 &   0.207 &   0.398 \nl
[Fe {\sc vii}]6087/[O {\sc i}]6300       &   0.014 &   0.071 &   0.136 \nl
[Fe {\sc x}]6374/[O {\sc iii}]5007       & --0.109 & --0.568 & --1.093 \nl
[Fe {\sc x}]6374/[S {\sc ii}]6717,6731   &   0.022 &   0.112 &   0.216 \nl
[Fe {\sc x}]6374/[O {\sc i}]6300         & --0.005 & --0.024 & --0.046 \nl
\enddata
\tablenotetext{a}{This value is the Circinus galaxy's one 
                  given by Oliva et al. (1994).}
\end{deluxetable}
%%%%%%%%%% TABLE 13 %%%%%%%%%%

%%%%%%%%%% TABLE 14 %%%%%%%%%%
\begin{deluxetable}{llcc}
\tablenum{14}
\tablewidth{0pt}
\tablecaption{Observed and Calculated High-Ionization
              Iron Emission-Line Ratios \label{tbl-14}}
\tablehead{
\colhead{} &
\colhead{Model Parameter} &
\colhead{[Fe {\sc x}]6374/[Fe {\sc vii}]6087} &
\colhead{[Fe {\sc xi}]7892/[Fe {\sc x}]6374} 
}
\startdata
\cutinhead{Observations\tablenotemark{a}}
NLS1    &                  & 0.936 $\pm$ 0.604 (9)  & 0.987 $\pm$ 0.504 (5) \nl
BLS1    &                  & 0.747 $\pm$ 0.391 (11) & 0.468 $\pm$ 0.135 (4) \nl
S1.5    &                  & 0.770 $\pm$ 0.884 (25) & 1.033 $\pm$ 0.622 (6) \nl
S2$^+$  &                  & 0.977 $\pm$ 1.239 (14) & 1.614 $\pm$ 0.358 (2) \nl
S2$^-$  &                  & 0.399 $\pm$ 0.174 (5)  & 0.598 $\pm$ 0.098 (2) \nl
Mean     &                  & 0.806 $\pm$ 0.863 (64) & 0.917 $\pm$ 0.563 (19) \nl
\cutinhead{Models}
FO86\tablenotemark{b} &                             & 0.10  & \nodata\nl
LOC\tablenotemark{c}  & Solar abundance             & 0.306 & 4.53   \nl
                 & Dusty abundance\tablenotemark{d} & 0.222 & 4.00   \nl
KF89\tablenotemark{e} & log $n_{e}$ = 0.0 cm$^{-3}$ & 0.959 & 0.557  \nl
                      & log $n_{e}$ = 0.5 cm$^{-3}$ & 1.299 & 0.581  \nl
                      & log $n_{e}$ = 1.0 cm$^{-3}$ & 1.622 & 0.581  \nl
VAC89\tablenotemark{f}& {\it V}$_{\rm shock}$ = 100 km s$^{-1}$& \nodata & \nodata \nl
                      & {\it V}$_{\rm shock}$ = 300 km s$^{-1}$& 1.408   & \nodata \nl
                      & {\it V}$_{\rm shock}$ = 500 km s$^{-1}$& 1.754   & \nodata \\
\enddata
\tablenotetext{a}{The number of objects including radio-loud galaxies are in parentheses.}
\tablenotetext{b}{Ferland \& Osterbrock (1986)}
\tablenotetext{c}{The locally optimally emitting cloud model 
                  proposed by Ferguson et al. (1997)}
\tablenotetext{d}{The abundance of the Orion nebula is assumed.}
\tablenotetext{e}{Korista \& Ferland (1989)}
\tablenotetext{f}{Viegas-Aldrovandi \& Contini (1989)}
\end{deluxetable}
%%%%%%%%%% TABLE 14 %%%%%%%%%%
\end{document}